\documentclass[a4paper,english,fleqn]{article}
\pdfoutput=1

\usepackage{amsmath}
\usepackage{amssymb}

\makeatletter

\pdfpageheight\paperheight
\pdfpagewidth\paperwidth

\usepackage[english]{babel}
\usepackage{authblk}
\usepackage{amsfonts}
\usepackage{amsthm}
\usepackage{color}
\usepackage{empheq}
\usepackage{graphicx}
\usepackage{float}
\usepackage{bbold}

\usepackage[colorlinks,linkcolor=blue,anchorcolor=black,citecolor=red]{hyperref}

\def\be{\begin{equation}}
\def\ee{\end{equation}}
\def\bc{\begin{center}}
\def\ec{\end{center}}

\newcommand{\N}{{\mathbb{N}}}
\newcommand{\R}{{\mathbb{R}}}
\newcommand{\E}{{\mathbb{E}}}
\renewcommand{\P}{{\mathbb{P}}}

\makeatother

\usepackage{babel}
\begin{document}

%
%
%
%

\title{Lower current large deviations for zero-range processes on a ring}

\author[*]{Paul Chleboun}
\author[**]{Stefan Grosskinsky}
\author[**]{Andrea Pizzoferrato}
\affil[*]{Department of Statistics, University of Oxford}
\affil[**]{Mathematics Institute, University of Warwick}

\renewcommand\Authands{ and }

%
%
%

\maketitle
%
%
%
%
\begin{abstract}
We study lower large deviations for the current of totally asymmetric zero-range processes on a ring with concave current-density relation. We use an approach by Jensen and Varadhan which has previously been applied to exclusion processes, to realize current fluctuations by travelling wave density profiles corresponding to non-entropic weak solutions of the hyperbolic scaling limit of the process. We further establish a dynamic transition, where large deviations of the current below a certain value are no longer typically attained by non-entropic weak solutions, but by condensed profiles, where a non-zero fraction of all the particles accumulates on a single fixed lattice site. This leads to a general characterization of the rate function, which is illustrated by providing detailed results for four generic examples of jump rates, including constant rates, decreasing rates, unbounded sublinear rates and asymptotically linear rates. Our results on the dynamic transition are supported by numerical simulations using a cloning algorithm.
\end{abstract}

%
%
%
%

\tableofcontents{}

%
%
%
%
\section{Introduction}

The large deviation behaviour of dynamic observables has been a topic of major recent research interest in driven diffusive systems. Most studies, as summarized in a recent review \cite{lazar1}, focus on the particle current as one of the most important characteristics of nonequilibrium systems in one dimension. In general, current fluctuations are studied from a microscopic or macroscopic point of view. 
For the first perspective, algebraic techniques are implemented to calculate eigenvalues and eigenvectors of an exponential tilted version of the generator of a stochastic lattice gas. In this way, the rate function of the large deviations of the current is calculated as a Legendre-Fenchel transform of the greatest eigenvalue of the tilted generator. These methods were successfully applied to the asymmetric simple exclusion process (ASEP) \cite{gorissen2012exact,Bodineau2006}, also in combination with the matrix product ansatz \cite{derrida2007non}, and to zero-range processes (ZRP) \cite{Harris2005,Harris2013,Hirschberg2015}. 
The statistics of the current and symmetry properties of the rate function can also be understood in the framework of the fluctuation theorem \cite{Lebowitz1998}. However, the symmetry relation stemming from the fluctuation theorem, also called Gallavotti-Cohen symmetry, breaks down in high current regimes for some condensing systems \cite{harris75breakdown,rakos2008range}. Almost all previous studies focus on open boundary conditions, with only few available for periodic boundary conditions \cite{popkov2010asep, tsobgni2016large}, where microscopic results are difficult to obtain due to temporal correlations \cite{indiansZRP07}.

From the macroscopic point of view, one of the most powerful frameworks introduced in recent years is the macroscopic fluctuation theory (MFT) (see \cite{Bertini2014a} and references therein), whose more general rigorous description is based on empirical flows \cite{3bigsfirst,3bigssecond}. This is able to provide, as a result of a variational principle, the time evolution of the most likely density profile which typically gives rise to a given fluctuation. It turns out that it can be hard to solve the variational problem and an expression for the density profiles has only been obtained for some specific models \cite{Bertini2014a,lazar1}. 

In general, macroscopic approaches rely on a hydrodynamic description of the process in terms of a mass conservation law. Lower current deviations, that is fluctuations of the current below its typical value, are usually realized by phase separated states for systems with concave flux function such as the exclusion process. These states can be described as weak solutions of the conservation law on a hydrodynamic level, while upper large deviations of the current are associated to hyperuniform states with long-range correlations \cite{jack2015hyperuniformity, karevski2016conformal}. The connection between hydrodynamics and large deviations is provided by the well-known concept of entropy production in weak solutions that exhibit shocks \cite{Smoller}. 
Using all possible entropy functionals, this can be used to identify a unique entropic solution to the hydrodynamic equation describing the typical behaviour. 
For non-entropic solutions the entropy production can provide the large deviation rate function for observing such a non-typical profile, if the correct thermodynamic entropy is used \cite{Varadhan2004}. This connection has been proved rigorously for the ASEP \cite{Jensen,Vilensky2008}, giving rise to the so-called Jensen-Varadhan theory. In \cite{DerriBodJV}, this has been applied heuristically to obtain a macroscopic derivation of the rate function for lower current deviations, which coincide with results based on exact microscopic computations and are in agreement with MFT predictions.

In this paper, we extend the Jensen-Varadhan approach to study lower current deviations for ZRPs which have a concave current-density relation. We focus on totally asymmetric dynamics with periodic boundary conditions, for which only few results exist so far. The ZRP was originally introduced in \cite{spitzer70} and it has simple stationary distributions of factorized form \cite{Andjel1982} which allow for a detailed stationary analysis. 
At the same time ZRPs can exhibit a condensation transition in homogeneous systems due to particle interactions when the density exceeds a critical value \cite{drouffe98,evansBrazil}. This has been studied in detail in recent years (see e.g. \cite{Evans2005,godreche,godreche2012condensation} and references therein), and has seen many applications \cite{eggerssand, macroZRP, trafficZRP}, as well as rigorous mathematical work (see e.g. \cite{Chleboun2014} and references therein). Here we focus on densities below the critical value, but we establish a dynamic transition for certain ZRPs where for sufficiently small currents the large deviations are dominated by condensed profiles rather than profiles arising from the Jensen-Varadhan approach. Our main result is a complete characterization of the rate function for lower current deviations for general totally asymmetric ZRPs with concave flux function.

The remainder of the paper is structured as follows. In Section \ref{sec:2}, we define stochastic lattice gases in terms of generators and we define current conditioning in the context of large deviation theory. We introduce four generic classes of ZRPs with concave flux function, which we will analyize throughout the paper using specific examples of jump rates. In Section \ref{sec:3} we present a general formulation of the Jensen-Varadhan approach for ZRPs, and compare corresponding cost functions for large deviation events to those of condensed states. 
Section \ref{sec:lddf} contains a detailed study of generic examples of ZRPs introduced in Section \ref{sec:2} which cover several cases of possible behaviour, two of which exhibit the dynamic transition.
\section{Definitions and Setting}\label{sec:2}
%
%
\subsection{TAZRP on a ring}
Consider a one-dimensional lattice $\Lambda$ with $\left|\Lambda\right| =L\in\mathbb{N}$ sites and  periodic boundary conditions, so that sites $L+1$ and $1$ coincide. Each site $x\in\Lambda$ can accommodate an integer number of particles $\eta_x\in\mathbb{N}$, 
and a configuration of the system is denoted by $\eta=\left(\eta_{1},\eta_{2},...,\eta_{L}\right)\in X_L$, where $X_L =\mathbb{N}^{\Lambda}$ is the configuration space. 
We focus on totally asymmetric zero-range processes (TAZRP), where particles only jump one site to the right with a rate $u :\N\to [0,\infty )$ that depends only on the occupation number of the departure site. The dynamics of the process can be described  by the generator
\begin{equation}\label{gen}
	\mathcal{L}f\left(\eta\right)=\sum_{x\in\Lambda}u\left(\eta_{x}\right)\left[f\left(\eta^{x,x+1}\right)-f\left(\eta\right)\right] \ ,
\end{equation}
for all test functions $f:X_L \to \mathbb{R}$. 
Since we consider only finite lattices there are no restrictions on the observable $f$, see \cite{Andjel1982} for details on infinite lattices. 
As usual, we denote by $\eta^{x,x+1}$ the configuration obtained from $\eta$ after a particle jumps from site $x$ to $x+1$, i.e. $\eta^{x,x+1}_y =\eta_y -\delta_{y,x} +\delta_{y,x+1}$. 
To avoid degeneracies and for later convenience we assume that the rates are in fact defined by a smooth function $u:\R\to [0,\infty )$ with
\begin{equation}\label{eq:transrule}
u\left(n\right)=0\mbox{ if and only if }n=0\quad\mbox{and}\quad u\left(n\right)>0\mbox{ for all }n>0\ .
\end{equation}
The process is irreducible on the state space $X_{L,N}\coloneqq\left\{ \eta\in X_L:\sum_{x\in\Lambda}\eta_{x}=N\right\}$ for each fixed $N\geq 0$, and the total number of particles is a conserved quantity under the dynamics.
We denote the process by $(\eta (t) :t\geq 0)$, with path space distribution $\P$ and the corresponding expectation by $\E$.
Most of our results will hold for general initial conditions and so it is not typically included in the notation. 
If we want to specify a certain initial configuration $\eta$ we will write $\P_\eta$ and $\E_\eta$. 

Under condition (\ref{eq:transrule}) 
it is known that the process admits stationary product measures, the so-called \emph{grand-canonical measures}, 
\begin{equation}\label{pm}
	\nu_{\phi}^{\Lambda}\left[d\eta\right]\coloneqq\prod_{x\in\Lambda}\nu_{\phi}\left(\eta_x\right)d\eta
\end{equation}
with a parameter $\phi\geq 0$, called the \textit{fugacity} \cite{spitzer70,Andjel1982}. 
The mass function of the single site marginal with respect to the counting measure $d\eta$ on $X_L$,  is given by
\begin{equation}
\nu_{\phi} \left(\eta_{x}\right)=\frac{1}{z\left(\phi\right)}w\left(\eta_{x}\right)\phi^{\eta_{x}}\ ,
\label{marginals}
\end{equation}
%
with stationary weights
\begin{equation}\label{weights}
	w\left(\eta_{x}\right)=\prod_{k=1}^{\eta_{x}}\frac{1}{u\left(k\right)} \quad\text{where}\quad w\left(0\right)=1\ ,
\end{equation}
and normalization
\begin{equation}\label{eq:partfct}
	z\left(\phi\right)=\sum_{n=0}^{\infty}w\left(n\right)\phi^{n} \ .
\end{equation}
$z(\phi)$ is also called the \textit{grand-canonical partition function}, and the measures $\nu_\phi$ exist for all $\phi\geq 0$ such that $z(\phi )<\infty$. We denote by $\phi_c \in (0,\infty ]$ the radius of convergence of $z(\phi )$, which we assume to be strictly positive. A convenient sufficient condition to ensure this, is that the jump rates are asymptotically bounded away from $0$, i.e. $\liminf_{k\to\infty} u(k)>0$ (see e.g.\ \cite{Chleboun2014}).

Under the grand-canonical measures the total particle number is random, and the fugacity parameter controls the average density
\begin{equation}\label{density}
	R\left(\phi\right)\coloneqq\left\langle\eta_{x}\right\rangle_\phi \coloneqq \sum_{n\in\N} \nu_\phi (n)n =\phi\,\partial_{\phi}\ln z\left(\phi\right)\ ,
\end{equation}
where we use the notation $\langle\cdot\rangle_\phi$ for expectations w.r.t.\ the distribution $\nu_\phi$. 
In general, $\ln z\left(\phi\right )$ is known to be a convex function, and $R\left(\phi\right )$ is striclty increasing in $\phi$ and continuous with $R\left(0\right) =0$ and largest value
\begin{equation}
\rho_c :=\lim_{\phi\nearrow\phi_c} R(\phi )\in (0,\infty ]\ .
\label{rhocdef}
\end{equation}
This is also called the critical density, and if finite, the system only has homogeneous stationary product measures with a bounded range of densities with $\nu_{\phi_c}$ being the maximal invariant measure. 
We denote the inverse of $R(\phi )$ by $\Phi\left(\rho\right)$.

Restricted to $X_{L,N}$ the unique stationary distribution is given by  
 conditioning the grand-canonical distribution to a fixed number of particles.
These are called the canonical stationary measures, they are independent of $\phi$ and are given by
\begin{equation}
\pi_{L,N}\left(\eta\right)\coloneqq\nu_{\phi}^{L}\left(\eta\left|X_{L,N}\right.\right) =\frac{\mathbb{1}_{X_{L,N}}(\eta)}{Z_{L,N}}\prod_{x\in\Lambda_{L}}w\left(\eta_x \right)\ ,
\end{equation}
where $Z_{L,N}\coloneqq\sum_{\eta\in X_{L,N}}\prod_{x} w\left(\eta_x\right)$ is the canonical partition function.
We denote  the average with respect to $\pi_{L,N}$ by $\langle \,\cdot\, \rangle_{L,N}$.

%
%
\subsection{Current large deviations}

For the TAZRP, the average stationary current w.r.t. to the canonical measure is defined as
\begin{equation}\label{cancurr}
J_{L,N} \coloneqq\langle u\rangle_{L,N},
\end{equation}
while under the grand-canonical measures we have
\begin{equation}\label{current}
	J\left(\rho\right)\coloneqq\left\langle u\right\rangle_{\Phi (\rho )}=\Phi\left(\rho\right)\ ,
\end{equation}
which is in fact given by the inverse of (\ref{density}), as a direct consequence of the form of the stationary weights (\ref{weights}). Due to the equivalence of ensembles (see e.g. \cite{Chleboun2014} and references therein), these two quantities are equivalent in the thermodynamic limit, i.e.\ for all $\rho <\rho_c$
\begin{equation}
J_{L,N} \to J(\rho )\quad\mbox{as }L,N\to\infty\quad\mbox{with }N/L\to\rho\ .
\label{equivalence}
\end{equation}
The (random) empirical current averaged over sites up to time $t>0$ is given by
\begin{equation}
\mathcal{J}^{L}\left(t\right)\coloneqq\frac{1}{L}\sum_{x}\mathcal{J}_{x,x+1}^{L}\left(t\right)
\label{emcu}
\end{equation}
where
\begin{equation}
\mathcal{J}_{x,x+1}^{L}\left(t\right)\coloneqq\frac{1}{t}\int_{0}^{t}\delta\left(1-\eta_{x}\left(s^{-}\right)+\eta_{x}\left(s\right)\right)ds
\end{equation}
is the current across the bond $x,x+1$ per unit time. 
For fixed $L$ and $N$ the ZRP is a finite-state, irreducible Markov chain on $X_{L,N}$, and a general approach in \cite{3bigsfirst, 3bigssecond} implies a large deviation principle (LDP) for the empirical current (\ref{emcu}) in the limit $t\to\infty$. 
The authors establish an LDP for general empirical densities and flows on path space, and the particle current is a continuous and in fact linear function of the empirical flow.
Then using the contraction principle (see e.g. \cite{Hollander,Touchette2009a}) and linearity they were able to show that the current $\mathcal{J}^{L} (t)$ satisfies an LDP with a convex rate function. 
We denote the associated rate function by $I^L$, and following the usual compact formulation for LDPs (see e.g.\ \cite{Touchette2009a}) on the level of logarithmic equivalence we have for all lower deviations $j\leq J(\rho )$
\begin{equation}
\mathbb{P}\left[\mathcal{J}^{L}\left(t\right)\leq j\right]\asymp e^{-tI^{L}\left(j\right)} \quad\mbox{as }t\to\infty\ .
\end{equation}
Based on results in \cite{BodADD,Bodineau2006} for the ASEP on a one-dimensional ring, our main result is a derivation of the rate function for diverging system size
\begin{equation}
I (j)=\lim_{L\to\infty} I^L (j)\ ,
\label{ratefu}
\end{equation}
for lower deviations $j\leq J(\rho )$. 
We focus on TAZRPs where
\begin{equation}
J \left(\rho \right)\quad\mbox{is a non-linear, concave, increasing function},
\label{jass}
\end{equation}
equivalently $R(\phi)$ is a non-linear convex increasing function of $\phi$. Linear functions would correspond to independent particles, which are not covered by our general approach, but are of course simple to treat and will be discussed later in Section \ref{sc:ALR}. 
Note that for all ZRPs, $J\left(\rho\right)$ and $R\left(\phi\right)$ are increasing, and so the only restriction is on the convexity. In addition to macroscopic arguments based on the Jensen-Varadhan approach for exclusion processes \cite{Jensen}, we also present simulation results based on the grand-canonical or tilted path ensemble 
\cite{Giardina2011, Harris2013, Chetrite2014}. This provides access to the scaled cumulant generating function defined as
\begin{equation}
\lambda^L\left(k\right)\coloneqq\lim_{t\to\infty}\frac{1}{t}\ln \mathbb{E}\left[e^{tk\mathcal{J}^{L}\left(t\right)}\right] .
\label{eq:momgen}
\end{equation}
Since the rate function is convex, it is then given by the Legendre-Fenchel transform
\begin{equation}
\label{eq:leg}
I^{L}\left(j\right)=\sup_{k\in\mathbb{R}}\left\{ kj-\lambda^L\left(k\right)\right\} \ .
\end{equation}
Since the current is a time-additive functional, we expect large deviations to be realized homogeneously in time, i.e. modulo a transient depending on the initial conditions, the function $s\mapsto\mathcal{J}^{L}\left(s\right)$ conditioned on $\mathcal{J}^{L}\left(t\right)\leq j$ is roughly constant and equal to $j$ for $s\leq t$. For a discussion of examples where conditioning does not lead to time-homogeneous behaviour see e.g. \cite{angeletti2016}.

In analogy to results for exclusion processes \cite{Bodineau2006}, we will see that if the system does not exhibit condensation ($\rho_c = \infty$) then typical realizations of lower current deviations for large $L$ are dominated by phase separated states which are non-entropic weak solutions of the hydrodynamic limit of the ZRP (see Section \ref{sec:hydroJV}) with two spatially separated regions at different densities. Since the phase boundaries move at non-zero speed we will refer to these as travelling wave profiles, which may exist only in a limited range of conditional currents. 
Outside this range, or for systems with finite critical density ($\rho_c < \infty$), condensed states may dominate the current large deviation, where a finite fraction of particles concentrates on a single, fixed lattice site.


%
%
\subsection{Generic examples}\label{subsec:genexemp}

In the following, we will discuss some examples of TAZRPs which obey \eqref{jass} and will be used throughout to illustrate our results. This includes models with bounded and unbounded jump rates.

The simplest example is given by constant jump rates
\begin{equation}
u\left(n\right)=1\quad\mbox{for all }n\geq 1\quad\mbox{and}\quad u(0)=0\ .
\label{rates_cr}
\end{equation}
In this case, the stationary measure $\nu_{\phi} (\eta_x )=\left(1-\phi\right)\phi^{\eta_x}$ is simply a geometric distribution, and the main quantities involved in the description of the process can be computed explicitly as
\begin{equation}
z\left(\phi\right)=\frac{1}{1-\phi}\ ,\quad R\left(\phi\right)=\frac{\phi}{1-\phi}\quad\mbox{and}\quad J\left(\rho\right)=\frac{\rho}{1+\rho}\ .
\label{eq:crrelations}
\end{equation}
Note that all densities $R(\phi )\geq 0$ are admissible, i.e. there exists a $\phi \geq 0$ such that $R(\phi )=\rho$, while the current $J(\rho )\in [0,1)$ due to the bounded jump rates. This process is equivalent to the TASEP (see e.g.\ \cite{warbook} or Appendix \ref{sec:mapZRPEP}) and its current fluctuations have been studied before \cite{Bodineau2006}, we simply include it for completeness.

The second example with bounded jump rates we will consider is given by
\begin{equation}\label{rates_cond}
u(0)=0\ ,\quad u\left(n\right)=1+\frac{b}{n}\quad\mbox{for all }n\geq 1\ ,\quad\text{with }b>0\ .
\end{equation}
This class of processes has been introduced in \cite{drouffe98,evansBrazil} and is known to exhibit a condensation phenomenon for $b>2$. It is easy to see that the stationary weights asymptotically decay as $w(n)\sim n^{-b}$, so that the stationary measures (\ref{pm}) exist for all $\phi\leq \phi_c =1$. This leads to a bounded range of admissible densities $R(\phi )\in [0,\rho_c ]$, with a finite critical density given by \cite{godreche,gss}
\begin{equation}
\rho_c =R(1)=\frac{1}{b-2}\ .
\label{rhoc}
\end{equation}
If conditioned on particle numbers $N\gg \rho_c L$ for large $L$, the system phase separates into a fluid phase, which is homogeneously distributed as $\nu_{\phi_c}$, and a condensed phase or condensate, where a finite fraction of $(\rho -\rho_c )L$ particles concentrates on a single lattice site (see e.g.\ \cite{gss,al1,Evans2005}). The interesting feature for this paper is that in addition to the density, also the range of admissible currents $j\leq J(\rho )$ by travelling wave profiles is bounded as explained in Section \ref{sc:condTAZRP}. The partition function $z(\phi )=\, _2 F_1 (1,1;1+b;\phi )\coloneqq \sum_{n=0}^\infty \frac{(1)_n (1)_n}{(1+b)_n} \frac{\phi^n}{n!}$ can be written in terms of hypergeometric functions $_2 F_1$ \cite{gss} using the Pochhammer symbol $(a)_n =\prod_{k=0}^{n-1} (a+k)$, which leads to similar expressions for for the convex function $R(\phi )$ and will be useful for numerical computations later.


We will also consider ZRPs with unbounded jump rates, for which it can be shown (see e.g. \cite{Chleboun2014}) that product measures exist for all $\phi \geq 0$, and all densities $\rho \geq 0$ are admissible. The first example we consider is
\begin{equation}\label{rates_ind}
u(0)=0\ ,\quad u\left(n\right)=n+d\quad\mbox{for all }n\geq 1\ ,\quad\text{with }d>0\ .
\end{equation}
Note that a rate $u(n)=n$ would correspond to independent particles jumping with rate $1$, leading to a linear current $J(\rho )=\rho$ and this degenerate case is not covered by our theory. Independent particles are easy to study with other tools, but they also arise as the limit $d\to 0$ of the above family of rates as we will discuss in Section \ref{sc:ALR}. The current behaves asymptotically as $J(\rho )\simeq u(\rho )=d+\rho$ for $\rho\to\infty$. Again, the main quantities can be computed explicitly in terms of known special functions as
\begin{equation}
z(\phi )=d e^\phi \phi^{-d} \big(\Gamma [d] - \Gamma [d, \phi ]\big)\ .
\label{zphigamma}
\end{equation}
where $\Gamma [d]$ and $\Gamma [d,\phi]$ are the complete and incomplete Euler gamma function, respectively. In particular, this implies that $R(\phi )=\phi \partial_\phi \ln z(\phi )$ is a convex function.

The second example with unbounded rates is given by sub-linearly diverging jump rates of the form
\begin{equation}\label{rates_sublin}
u\left(n\right)=\frac{\left[\left(n+1\right)^{\gamma}-1\right]}{\gamma}\ ,\quad\text{with }\gamma\in (0,1)\ .
\end{equation}
Rather than $n^\gamma$ we use this regularized functional form for the rates, since $u'(0)=1$ and it converges uniformly to $u(n)=\ln (n+1)$ as $\gamma\to 0$ which can be studied as a limiting case. Again, all densities are admissible with $\rho_c =\infty$. We are not aware of known special functions that lead to exact expressions for the partition function $z$ to simplify the numerics in this case. $J(\rho )$ turns out to be concave for all $\rho\geq 0$ and behaves asymptotically as $J(\rho )\simeq u(\rho )\simeq (1+\rho )^\gamma /\gamma$ as $\rho \to \infty$.

%
%
\subsection{Hydrodynamics and the Jensen-Varadhan functional}\label{sec:hydroJV}


It is well known that the large-scale dynamics of the asymmetric ZRP in hyperbolic scaling $y=x/L,\ \tau =t/L$ is described in a hydrodynamic limit by the conservation law for the density field $\rho\left(y,\tau\right)=\mathbb{E}\left[\eta_{yL}\left(\tau L\right)\right]$,
\begin{equation}\label{pde}
		\frac{\partial}{\partial \tau}\rho \left(y,\tau\right)+\frac{\partial}{\partial y}J\left(\rho\left(y,\tau\right)\right)=0\quad y\in\mathbb{T},\ \tau\geq 0\ .
\end{equation}
Here $\mathbb{T}$ denotes the unit torus, which arises due to periodic boundary conditions. 
This has been proved rigorously for non-decreasing jump rates using coupling techniques (see e.g. \cite{Landim} and references therein). For ZRPs with decreasing rates as in (\ref{rates_cond}), there are recent results for symmetric systems \cite{stamatakis} for sub-critical densities, but the description by (\ref{pde}) is believed to hold also for asymmetric systems \cite{Schutz2007}. For a given initial condition $\rho (y,0)$ the above equation can be solved using the method of characteristics \cite{laxbook}, which are curves $(y(\tau ),\tau )$ along which the solution is constant, i.e. $\rho (y(\tau ),\tau )=\rho (y(0),0)$. It is easy to see that for conservation laws of the form \eqref{pde} characteristics are in fact straight lines with characteristic speed $J'(\rho (y(0),0))$. 
Depending on the initial conditions characteristics can intersect, leading to the occurence of shocks and non-differentiable solutions even from smooth initial data, which are described by the concept of weak solutions which satisfy an integrated version of \eqref{pde} (see e.g.\ \cite{Smoller}, Section 15).
Without further restrictions, weak solutions are not unique and selection criteria have to be imposed to single out the physically relevant ones. Due to convergence of standard discretization schemes (see \cite{Smoller} for details), it turns out that it is sufficient to understand the solution of \eqref{pde} for the so-called Riemann problem with
\begin{equation}
\rho (y,0)=\left\{\begin{array}{cc} \rho_l &,\ y<0\\ \rho_r &,\ y\geq 0\end{array}\right.\quad\mbox{formulated for }y\in\R\ .
\label{riemann}
\end{equation}
If characteristics collide, a stable shock emerges with speed given by
%
%
\begin{equation}
v_{s} (\rho_l ,\rho_r) =\frac{J\left(\rho_r\right)-J\left(\rho_l\right)}{\rho_r-\rho_l}\ ,
\label{speed}
\end{equation}
which can be derived from the conservation of mass. 
The characteristic speeds for stable shocks fulfill
\begin{equation}
J'(\rho_l )>v_s (\rho_l ,\rho_r )>J'(\rho_r )\ .
\end{equation}
If characteristics drift apart, the solution is given by a rarefaction fan, which is a travelling wave solution that interpolates between the two densities $\rho_l$ and $\rho_r$. 
For the concave flux functions we consider, this implies that up shocks with $\rho_l <\rho_r$ are stable, while down shocks desolve in a rarefaction fan. 

An equivalent criterion to determine the uniqueness of weak solutions under general assumptions was developed by Kruzkov (see e.g.\ \cite{laxbook,Smoller}). 
Consider a regular convex function $h\left( \rho\right)$, called \textit{entropy}, with corresponding \textit{entropy flux} $g\left( \rho\right)$ such that
\begin{equation}
g'\left(\rho\right)=J'\left(\rho\right)h'\left(\rho\right).
 \label{eflux}
 \end{equation}
To select the physically relevant weak solution $\rho$ of (\ref{pde}) one requires that for all entropy-entropy flux pairs, again in a weak sense, 
\begin{equation}
		\frac{\partial}{\partial \tau}h(\rho \left(y,\tau\right) )+\frac{\partial}{\partial y}g\left(\rho\left(y,\tau\right)\right)\geq 0\ ,
\label{kruzkov}
\end{equation}
and with this additional constraint such entropy solutions are uniquely determined. Note that for smooth solutions equality holds in (\ref{kruzkov}) and entropy is a conserved quantity. 
Entropy is not conserved for shock solutions, and the inequality constraint ensures that entropy is produced across a shock, corresponding to the concept of information being irreversibly lost when characteristics collide. For a single shock with $\rho_l <\rho_r$, travelling with speed $v_s$ (\ref{speed}), integrating (\ref{kruzkov}) over space yields that the entropy production rate across the shock is given by 
\begin{equation}
\mathcal{F}\left(\rho_l ,\rho_r \right) \coloneqq g\left(\rho_l \right)-g\left(\rho_r \right)-\frac{J\left(\rho_r\right)-J\left(\rho_l\right)}{\rho_r-\rho_l}\left(h\left(\rho_l\right)-h\left(\rho_r\right)\right)\ .
\label{entprod}
\end{equation}
For stochastic particle systems with stationary product measures of the form \eqref{pm}, the thermodynamic entropy plays a special role, which is given by the Legendre transform of the pressure $\ln z(\phi )$ via
\begin{equation}\label{eq:entrflux}
h\left(\rho\right)=\rho\ln\Phi\left(\rho\right)-\ln z\left(\Phi\left(\rho\right)\right)\ .
\end{equation}
This is also equal to the relative entropy density $\frac{1}{L} H(\nu_\phi^L ,w^L )$ of the grand-canonical measures w.r.t.\ the stationary weights $w^L$ (\ref{weights}), see \cite{gss,Varadhan2004,paulthesis,gc2015} for a general discussion. 
Using this entropy for the asymmetric exclusion process, it was shown in \cite{Jensen} and \cite{Vilensky2008} that the large deviation rate function to observe a non-entropic weak solution over a fixed macroscopic time interval $[0,\tau ]$ in the limit $L\to\infty$ is given by the accumulated negative part of the entropy production. So the reduction in entropy for non-entropic solutions provides a purely macroscopic quantification of how unlikely they are to be observed in the underlying stochastic model under hyperbolic scaling.

This result has been applied in \cite{Bodineau2006} heuristically in a different scaling. For fixed, large system size $L$, lower current deviations for the asymmetric exclusion process on a ring are realized by phase separated travelling wave step profiles with two densities $\rho_1 <\rho_2$, which are uniquely determined by the total mass and conditional current. The probabilistic cost to realize such a profile does not depend on system size since only the non-entropic down shock has to be stabilized. This cost is equal to the entropy production across the reversed stable shock given by $\mathcal{F} (\rho_1 ,\rho_2 )$, which is also equal to $-\mathcal{F} (\rho_2 ,\rho_1 )$ by obvious symmetry in \eqref{entprod}.

%
%
%
%

\section{General results}\label{sec:3}

Even though they are only proved for the asymmetric exclusion process, the results in \cite{Jensen,Varadhan2004,Vilensky2008} depend only on the hyperbolic scaling limit and are of a general nature that can, at least heuristically, be applied directly to other particle systems. 
Therefore we assume that the same formalism used for the exclusion process in \cite{Bodineau2006} applies to the ZRPs we consider here, since we assume that they also have concave flux functions $J(\rho )$. 

Below we described two efficient strategies for the process to realise a large deviation of the current $\mathcal{J}^L(t) \leq j < J(\rho)$.
The first is by travelling wave profiles, for which we can estimate the large deviation cost of realising a current $j$ using a Jensen-Varadhan approach, similar to that used in \cite{Bodineau2006} for the exclusion process. We denote this cost by $E_{\rm tw}(j)$ (see \eqref{twcost}).
Secondly, if the process can exhibit condensation under the stationary measures (i.e. $\rho_c < \infty$) we will see that such a large deviation in the current are sometimes  more efficiently realised by condensed states.
We denote the large deviation cost associated with realising a current $j< J(\rho)$ by a condensed state by $E_c(j)$ (see \eqref{eq:condcostlim}).
Our main result is that for any TAZRP with concave flux function the large deviation rate function \eqref{ratefu} in the limit $L\to\infty$ is given by 
\begin{equation}
I(j)=E_{tw}(j) \quad\mbox{for all }j<J(\rho )\,, \quad\mbox{ if } \rho_c=\infty\, ,
\label{mainres1}
\end{equation}
and is given by the lower convex hull
\begin{equation}
I(j)=\mathrm{\underline{conv}}\big\{ E_{tw} , E_c \big\} (j)\quad\mbox{for all }j<J(\rho )\,, \quad\mbox{ if } \rho_c<\infty\ .
\label{mainres2}
\end{equation}
This constitutes a dynamical phase transition, where the realization of current large deviations switches from travelling wave to condensed profiles for low enough values of $j$. Details on applying this to different examples and finite-size corrections for large $L$ will be discussed in Section \ref{sec:lddf}, in the following we provide definitions and general results for travelling wave and condensed profiles.

%
%
\subsection{Travelling wave profiles}

\begin{figure}[t]
\begin{center}
\mbox{\includegraphics[width=0.47\textwidth]{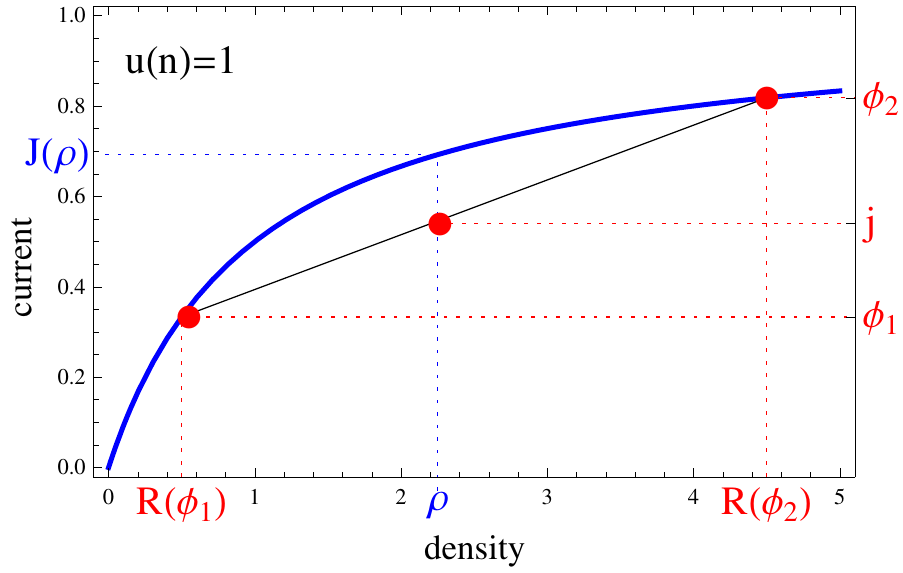}\quad\includegraphics[width=0.54\textwidth]{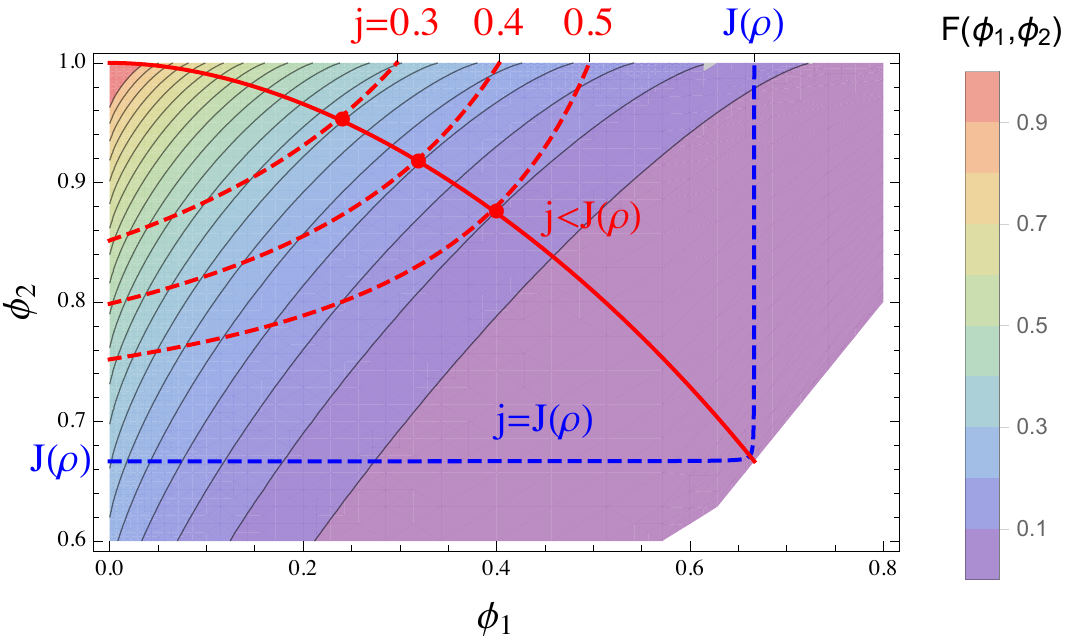}
}
\end{center}
\caption{\label{fig:illu}
The two plots feature the constant rate TAZRP \eqref{rates_cr}. (Left) The blue line depicts the current-density relation for the constant rate ZRP \eqref{rates_cr}, while the intersecting black line is obtained from the consistency relations \eqref{jcond} and \eqref{rcond}, varying the volume fraction $x$ between $0$ and $1$. For a fixed density $\rho$ each admissible pair $(\phi_1 ,\phi_2 )$ corresponds to a current $j<J(\rho )$. (Right) Contour plot of the Jensen-Varadhan functional \eqref{jv} is shown together with the constraint curves \eqref{cond} (red dashed lines), which are plotted for several values of $j<J(\rho )$. The blue dashed line is the limiting constraint line for $j\to J\left(\rho\right)$. The full red dots correspond to the minimizers of \eqref{eq:minsystem}. The union of all the optimal points is represented as a full red line.
}
\end{figure}

\begin{figure}[t]
\begin{center}
\mbox{\includegraphics[width=0.48\textwidth]{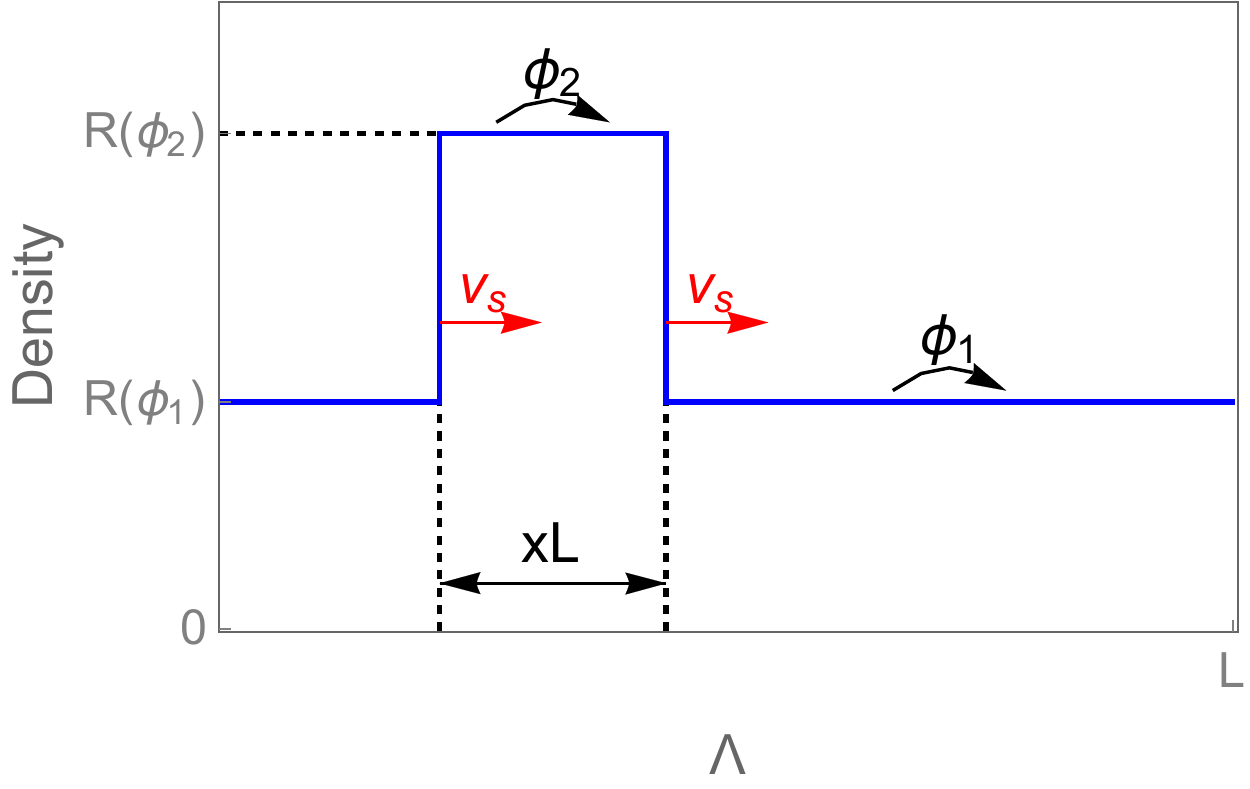}\quad\includegraphics[width=0.50\textwidth]{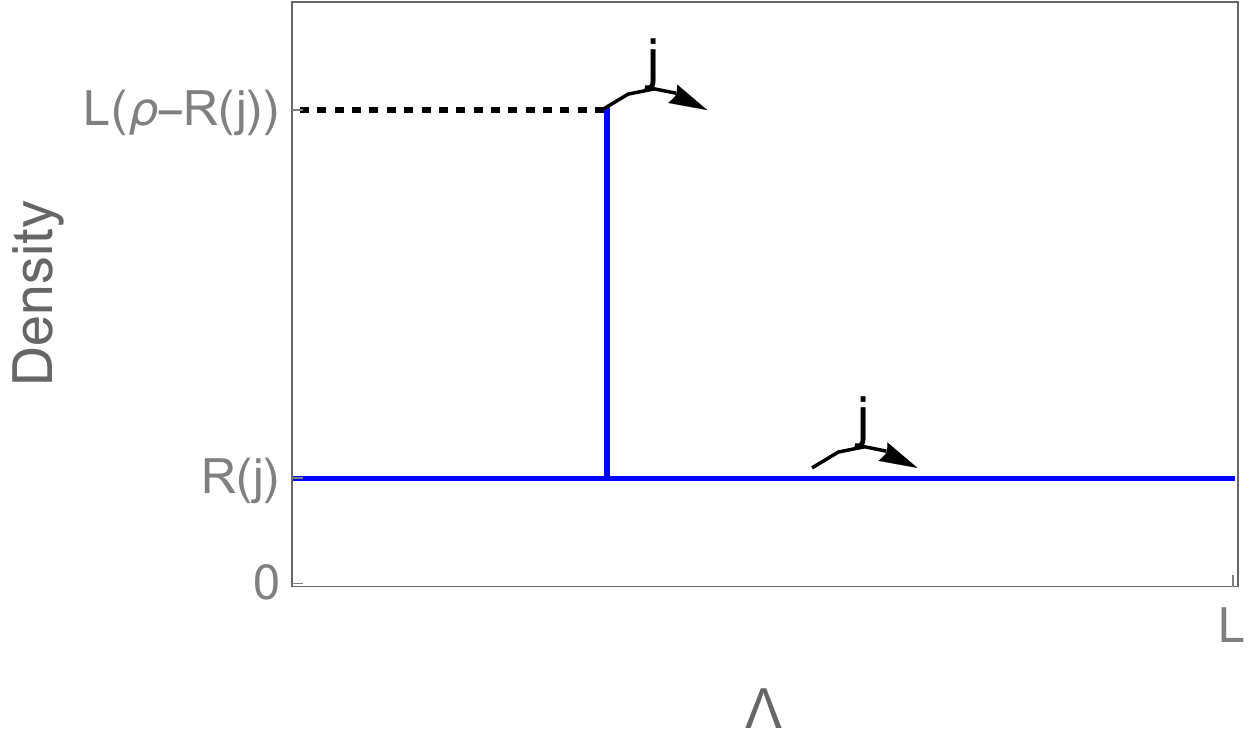}}
\end{center}
  \caption{\label{fig:profiles} Illustrations of phase separated profiles on the lattice $\Lambda$ with periodic boundary conditions. (Left) A traveling wave profile with high density region at density $R(\phi_1)$ and low density region at $R(\phi_2)$ satisfying \eqref{jcond} and \eqref{rcond}. The profile moves to the right with shock speed $v_s$ given by \eqref{eq:speedbis}. (Right) Condensed state profile with density of the fluid phase given by $R\left(j\right)$ and a fixed condensate of typical size $L\left(\rho-R\left(j\right)\right)$.
  }
\end{figure}

Travelling wave profiles are characterized by pairs of fugacities (or currents) $\phi_1 \leq j< J(\rho )\leq \phi_2$ under the constraints of fixed total density $\rho$ and total current $j<J(\rho )$. 
These constraints are characterized by
\begin{align}
j&=\left(1-x\right)\phi_1 +x\phi_2 \label{jcond}\\
\rho &=\left(1-x\right)R\left(\phi_{1}\right)+xR\left(\phi_{2}\right)\ ,\label{rcond}
\end{align}
where $x\in [0,1]$ parametrizes the volume fraction of the high density $\phi_2$ phase.
Since $\phi_1 <\phi_2$, by eliminating the variable
\begin{equation}\label{eq:hdensfrac}
x=\frac{j-\phi_{1}}{\phi_{2}-\phi_{1}},
\end{equation}
the constraints \eqref{jcond} and \eqref{rcond} can be re-written as
\begin{equation}
G\left(\phi_{1},\phi_{2}\right)\coloneqq\frac{\rho\left(\phi_{2}-\phi_{1}\right)-\phi_{2}R\left(\phi_{1}\right)+\phi_{1}R\left(\phi_{2}\right)}{R\left(\phi_{2}\right)-R\left(\phi_{1}\right)}=j,
\label{cond}
\end{equation}
which implicitly defines a one-dimensional subset of admissible fugacity pairs $(\phi_1 ,\phi_2 )$ explained in detail in Section \ref{sec:lddf}.
In Figure \ref{fig:illu} (left) all relevant quantities are illustrated for the constant rate ZRP, and Figure \ref{fig:profiles} (left) shows an illustration of a travelling wave profile.

The large deviation cost associated with such as traveling wave profile can be determined in terms of the thermodynamic entropy \eqref{eq:entrflux}.
Since the stationary current for the TAZRP is simply given by $J(\rho )=\Phi (\rho )$, it is easy to see that the corresponding entropy flux that fulfills (\ref{eflux}) is 
\begin{equation}
g(\rho )=\Phi (\rho )\big( \ln\Phi (\rho )-1\big)\ .
\label{zrpeflux}
\end{equation}
With the shock speed $v_s =\frac{\Phi(\rho_2 )-\Phi (\rho_1 )}{\rho_2 -\rho_1}$ the Jensen-Varadhan functional for a single shock (\ref{entprod}), which gives the large deviation cost,  can be written conveniently as a function of fugacities $\phi_i =\Phi (\rho_i )$ for a general ZRP, 
\begin{align}
F\left(\phi_{1},\phi_{2}\right)&\coloneqq \mathcal F\left(R\left(\phi_1\right),R\left(\phi_2\right)\right)=- \mathcal F\left(R\left(\phi_2\right),R\left(\phi_1\right)\right)\nonumber\\
&=g\big( R(\phi_1 )\big)-g\big( R(\phi_2 )\big) -v_s \Big[ h\big( R(\phi_1 )\big) -h\big( R(\phi_2 )\big)\Big]\nonumber\\
&=\Big[\left(\phi_{1}\ln\phi_{1}-\phi_{1}\right)-\left(\phi_{2}\ln\phi_{2}-\phi_{2}\right)\Big] -\Big[\frac{\phi_{2}-\phi_{1}}{R\left(\phi_{2}\right)-R\left(\phi_{1}\right)}\Big]\times\nonumber\\
&\quad \times\Big[\left(R\left(\phi_{1}\right)\ln\phi_{1}-\ln z\left(\phi_{1}\right)\right)-\left(R\left(\phi_{2}\right)\ln\phi_{2}-\ln z\left(\phi_{2}\right)\right)\Big]\ .
\label{jv}
\end{align}
The partition function $z(\phi )$ and density $R(\phi )=\phi\partial_\phi \ln z(\phi )$ can be computed (often explicitly) without the need of inverse functions, and current or fugacity are therefore more suitable variables than densities for ZRP. 

Important general properties of (\ref{jv}) are the following. $F(\phi_1 ,\phi_2 )$ is decreasing in $\phi_1$ and increasing in $\phi_2$, and it is anti-symmetric, i.e. $F(\phi_1 ,\phi_2 )=-F(\phi_2 ,\phi_1 )$. 
Therefore $F(\phi ,\phi)=0$, which corresponds to $0$ cost for vanishing step size, and it is positive for $\phi_2>\phi_1$. 
In all examples we have studied $F$ is also convex and has concave level lines, but we are not able to show this in general. In our examples, $F$ is also a smooth function on its domain of definition which is either $[0,\phi_c )^2$ or $[0,\phi_c ]^2$ in case of a condensing system with $\phi_c <\infty$. This is always the case as long as $\ln z$ is smooth.

Due to concavity of the flux function $J(\rho )$, the above profiles actually realize lower current deviations as is illustrated in Figure \ref{fig:illu} (left). 
We fix a density $\rho >0$ with an associated typical stationary current $J(\rho )$, and condition on a current $j< J(\rho )$. If the system has a finite critical density $\rho_c <\infty$, we also require $\rho <\rho_c$.
The rate function of the exponential cost to realize a travelling wave profile is then given by minimizing \eqref{jv} subject to the constraint \eqref{cond}, that is
\begin{equation}
\label{twcost}
E_{tw} (j)\coloneqq\inf\big\{ F(\phi_1 ,\phi_2 )\, :\, G\left(\phi_{1},\phi_{2}\right)=j\big\}\in [0,\infty ]\ .
\end{equation}
Depending on the regularity of $F$ and $G$ in a given example, the minimizer in \eqref{twcost} is often a local minimizer in the interior of the domain and can be found as a solution to the following system of equations
\begin{equation}
\left\{ \begin{array}{c}
\partial_{1}F\left(\phi_{1},\phi_{2}\right)\partial_{2}G\left(\phi_{1},\phi_{2}\right)-\partial_{2}F\left(\phi_{1},\phi_{2}\right)\partial_{1}G\left(\phi_{1},\phi_{2}\right)=0\\
G\left(\phi_{1},\phi_{2}\right)=j
\end{array}\right.\ .
\label{eq:minsystem}
\end{equation}
In general, it is not clear if there exists a unique minimizer in (\ref{twcost}) or whether it is a local or a boundary minimum, and it is not possible to get explicit expressions. 
We will see later in Section \ref{sec:lddf} that the infimum is usually unique, but that in some cases the constraint \eqref{cond} cannot be fulfilled and there are no travelling wave profiles, resulting in the cost in \eqref{twcost} being equal to $\inf \emptyset =\infty$. Travelling wave profiles with more than one up and one down step are more costly than the simple one shown in Figure \ref{fig:profiles} (left) and do not contribute to typical large deviation events.\\

\begin{figure}
\begin{center}
\mbox{\includegraphics[width=0.48\textwidth]{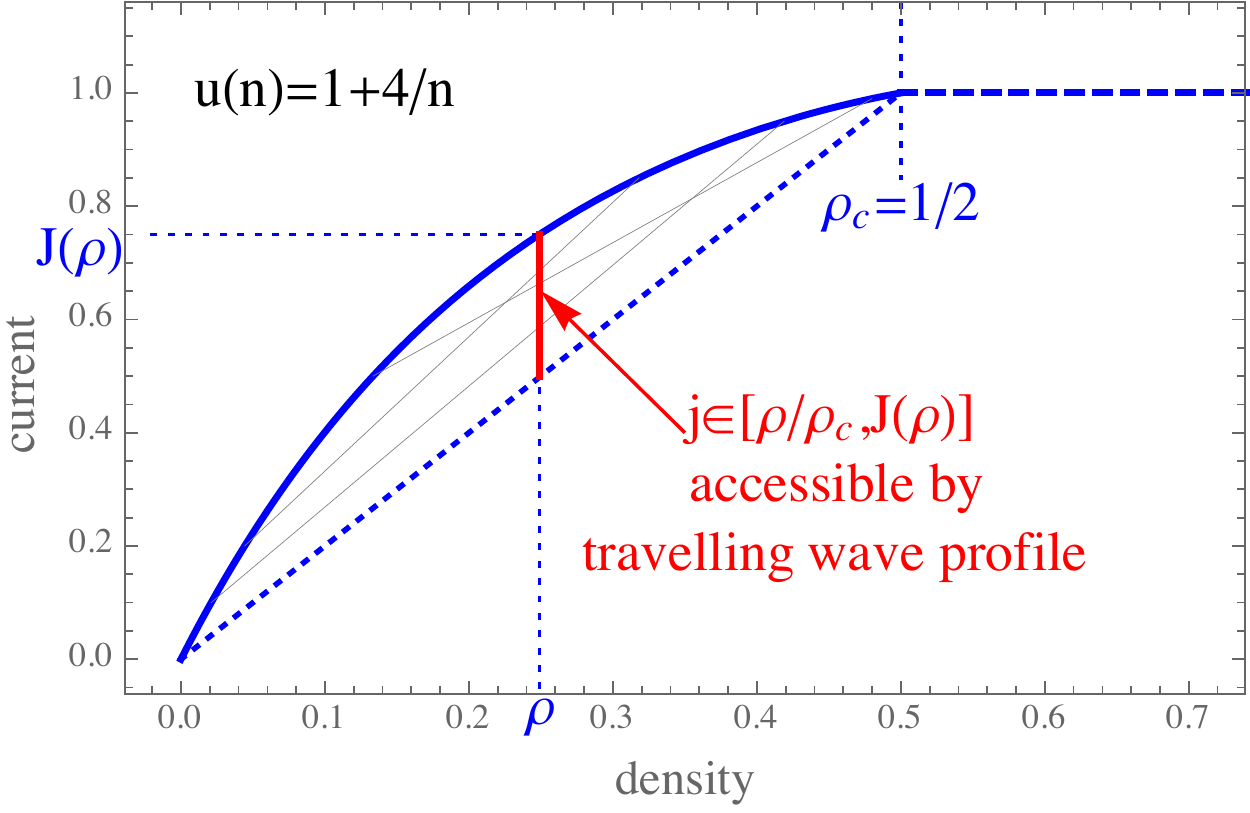}\quad\includegraphics[width=0.48\textwidth]{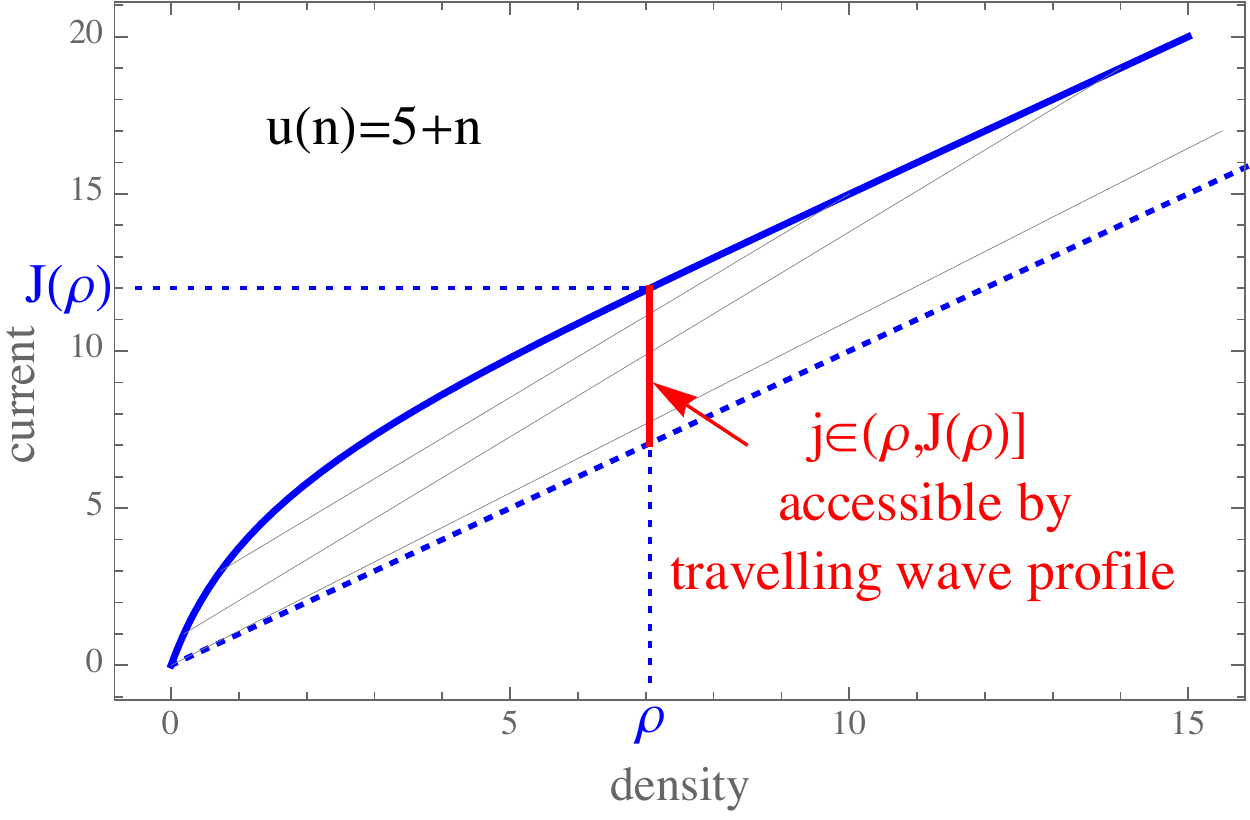}
}
\end{center}
\caption{\label{fig:brange}
Restricted range of currents $j\leq J(\rho )$ which are admissible by travelling wave profiles for a condensing process with rates (\ref{rates_cond}) (left), and for asymptotically linear rates (\ref{rates_ind}) (right). The grey lines indicate examples of admissible pairs $(\phi_1 ,\phi_2 )$ as in Figure \ref{fig:illu} (left).
}
\end{figure}

\noindent\textbf{Properties of the travelling wave profile.} For the constant rate example illustrated in Figure \ref{fig:illu}, picking $\phi_1 =0$, it is clear that all currents $0\leq j\leq J(\rho )$ are admissible for the constraint \eqref{cond} $G\left(0,\phi_2\right)=\rho\frac{\phi_2}{R\left(\phi_2\right)}=0$, since $\phi_2 /R(\phi_2 )=1-\phi_2 \to 0$ as $\phi_2\to 1$. 
As is illustrated in Figure \ref{fig:brange}, the smallest current $j$ admissible by travelling wave profiles is in general given by
\begin{equation}
j_{min}=\rho\lim_{\phi_2\nearrow\phi_c} \frac{\phi_2}{R(\phi_2 )}\ ,
\label{eq:jmin}
\end{equation}
where $\phi_c$ could be finite or infinite. A bounded range of admissible currents $j$ is possible due to a bounded range of densities in condensing systems (e.g.\ with rates (\ref{rates_cond})), where $j_{min}=\phi_c \frac{\rho}{\rho_c}$, or if $R(\phi )$ is asymptotically linear, as is the case for the system with rates (\ref{rates_ind}), where $j_{min}=\rho$.

It is clear from the illustration in Figure \ref{fig:illu} (left), and the fact that $\phi /R(\phi )$ is decreasing as a consequence of (\ref{jass}), that for given $\rho$ and admissible $j$, $\phi_2$ is uniquely determined by $\phi_1$. 
Therefore, for any admissible $j$ with $\phi_1\leq j\leq J(\rho )$ the solution of the constraint (\ref{cond}) implicitly defines a function
\begin{equation}\label{eq:consHL}
\bar\phi_2 (\phi_1 )\text{ such that } G(\phi_1,\bar\phi_2(\phi_1))=j\,,
\end{equation}
shown by dashed red lines in Figure \ref{fig:illu} (right).
$\bar\phi_2 (\phi_1 )$ is strictly increasing in $\phi_1$ and 
since $R(\phi )$ and its inverse are also continuous, $\bar\phi_2 (\phi_1 )$ is in fact a continuous increasing function for all $\phi_1 \in [0,j)$.
Actually, this domain is bounded above by a value strictly smaller than $j$ for systems with $j_{min} >0$, and for non-accessible currents $j<j_{min}$ the function \eqref{eq:consHL} is not defined. This applies to the examples in Sections \ref{sc:ALR} and \ref{sc:condTAZRP} and is discussed there in detail. 
At the left boundary for $\phi_1 =0$ the value of $\bar\phi_2(0)>0$ is the positive solution to
\begin{equation}
\rho\phi_2 =jR(\phi_2 ) \, ,
\end{equation}
which exists for all admissible $j<J(\rho )$ and is easily constructed graphically (see Figure \ref{fig:illu} (left)). We further note that the high density volume fraction $x$ \eqref{eq:hdensfrac} as well as the speed of profile 
\begin{equation}\label{eq:speedbis}
v_s =(\phi_2 -\phi_1)/(R(\phi_2 )-R(\phi_1 ))
\end{equation}
are decreasing with increasing $\phi_1$, and in systems with $j_{min} =0$ both vanish as $\phi_1 \to j$.

For all the examples we studied it further turns out that $\bar\phi_2(\phi_1)$ is convex, and with convexity of $F(\phi_1 ,\phi_2 )$ and resulting concave level lines, this leads to a unique minimum of the cost $F$ along the curve $(\phi_1,\bar\phi_2(\phi_1))$ as is illustrated in Figure \ref{fig:illu} (right) for the constant rate process. 
This minimum could be located inside the domain of definition, or located at the boundary $\phi_1 =0$ or $\phi_2 =\phi_c$ in the case $\phi_c <\infty$. 
The location of minima for different $j<J(\rho )$ is shown by a full red line in Figure \ref{fig:illu} (right). 
For the typical current $j=J(\rho )$ no condition on the system is imposed and the optimal pair is given by $\phi_1 =\phi_2 =J(\rho )$. 

Since we assume non-linearity and concavity of the function $J(\rho )$, it is clear from Figure \ref{fig:brange} that $j_{min} <J(\rho )$ and there are currents at least close to the typical one which are admissible by travelling wave profiles. 
Furthermore, due to smoothness of the constraint curve \eqref{cond} and the Jensen-Varadhan functional \eqref{jv}, and due to anti-symmetry of the latter, the travelling wave cost function \eqref{twcost} is continuous and $E_{tw} (J(\rho ))=0$ at the typical value for the current. 
Therefore $E_{tw} (j)$ itself is a proper rate function for the current, and in many cases $I(j)=E_{tw} (j)$.


\subsection{Condensed states}\label{sc:condensedstates}

A given current $j < J\left(\rho\right)$ can also be realized by the bulk of the system taking density $R(j)$ and all the excess mass $(\rho-R(j))L$ being located on one single (fixed) lattice site.
In general, when conditioning on a low current $j$, a stable condensed state is obtained when the current out of the condensate  matches the current $j<J(\rho )$ in the bulk phase of the system.
The condensate acts as a boundary reservoir, the exit rate of which has to be slowed down from a value of order $u\big( (\rho -R(j))L\big)$ to $j$, to assure the right incoming current into the bulk. 
Then the cost to maintain a stable condensate corresponds to the cost of slowing down a Poisson process across one bond (see e.g.\ \cite{Harris2005})
\begin{equation}
E^L_c (j)=u\big( (\rho -R(j))L\big) -j+j\ln \frac{j}{u\big( (\rho -R(j))L\big)}\ .
\label{condcost}
\end{equation}
This is not exact, since we simply replaced the argument of the rates $u(n)$ by an average value, but with our regularity assumptions \eqref{eq:transrule} on $u$ this is correct to leading order in $L$. Condensed phase separated profiles are illustrated in Figure \ref{fig:profiles}. Note that opposed to travelling wave profiles, the range of admissible currents for condesed states is always given by the full interval $\left[0,J\left(\rho\right)\right)$.\\

For unbounded rates $u$, $E^L_c (j)$ diverges as $L\to\infty$ of order $u\big( (\rho -R(j))L\big)$. 
However, travelling wave profiles always yield costs $E_{tw} (j)$ which are independent of the system size $L$ (see \eqref{twcost}) for $j_{\rm min} < j < J(\rho)$. 
For such systems the current rate function \eqref{mainres2} is therefore given by $I(j)=E_{tw} (j)$ for all $j>j_{min}$, and condensed profiles may only contribute in systems with bounded jump rates or if $j_{min} >0$ in which case not all currents are admissible by travelling wave profiles.
An example of the latter is given by asymptotically linear jump rates \eqref{rates_ind}, which is discussed in detail in Section \ref{sc:ALR}. \\

If $u$ is bounded and has a limit, we have $\phi_c =\lim_{k\to\infty} u(k)<\infty$ and for diverging system size the condensed cost converges to a finite value
\begin{equation}
E^L_c (j) \to E_c (j)=\phi_c -j +j\ln\frac{j}{\phi_c}\quad\mbox{as }L\to\infty\quad \textrm{if } \phi_c < \infty\ .
\label{eq:condcostlim}
\end{equation}
Examples of bounded jump rates, in particular the cases of constant rate and condensing ZRP are discussed below in Sections \ref{sc:CRTAZRP} and \ref{sc:condTAZRP}. 
Note that the expressions \eqref{condcost} and \eqref{eq:condcostlim} only apply for $j<J(\rho )$, and that $\lim_{j\to J(\rho )} E_c (j)>0$ does not vanish when approaching the typical current. In fact $E_c^L (J(\rho ))$ and $E_c (J(\rho ))$ are not well defined and depend on details of the limiting sequences involved in (\ref{eq:condcostlim}), so the condensed cost itself is not a valid large deviation rate function. 
However, we have seen above that travelling wave profiles are always admissible for currents $j$ just below $J(\rho )$ and $E_{tw} (J(\rho ))=0$. Therefore the rate function is always dominated by travelling wave profiles for $j$ sufficiently close to $J(\rho )$, and condensed profiles can only be relevant for lower values of $j$ where the description in \eqref{condcost} and \eqref{eq:condcostlim} is valid. \\

If the jump rates are bounded but $\rho_c = \infty$, that is the system does not exhibit condensation under the stationary measures for any density, we will now show that condensed profiles are always less likely than travelling wave profiles.
With bounded jump rates we have $\phi_c < \infty$ and  $\frac{R\left(\phi\right) }{\phi}\to\infty$ as $\phi\to\phi_c$.
This implies that $j_{min}= 0$ from (\ref{eq:jmin}), and includes for example the constant rate case. 
In order to compare condensed and travelling wave profiles, we fix the size of the high density phase to be $x=\frac{1}{L}$. Together with $j$ and $\rho$ this fixes a particular pair $\left(\phi_1^c,\phi_2^c\right)$ on the constraint curve \eqref{eq:consHL} which does not necessarily minimize \eqref{jv}. From the phase separation conditions (\ref{jcond}) and (\ref{rcond}), we have
\begin{equation}
x=\frac{1}{L}=\frac{j-\phi^c_1}{\phi^c_2-\phi^c_1}\quad\mbox{and}\quad R\left(\phi^c_2 \right)= L\rho -\left(L-1\right)R\left(\phi^c_1\right)\ .
\end{equation}
In the limit $L\to\infty$ this implies 
\begin{equation}
\phi^c_1 \simeq j\quad\mbox{with}\quad R\left(\phi^c_2 \right)\simeq L\left(\rho-R\left(j\right)\right)
\label{eq:denslim}
\end{equation}
and from (\ref{current})
\begin{equation}
\phi^c_2\simeq J\left(L\left(\rho-R\left(j\right)\right) \right)\to\phi_c.
\label{eq:phi2lim}
\end{equation}
The cost of such a travelling wave profile then satisfies
\begin{equation}
F(\phi^c_1 ,\phi^c_2)\to \phi_c -j +j\ln\frac{j}{\phi_c} =E_c (j)\quad\mbox{as }L\to\infty ,
\label{limcost}
\end{equation}
where we have used that $\ln z(\phi_2 )/R(\phi_2 )\to 0$ as $\phi \to \phi_c$ (see Lemma in Appendix \ref{sec:lemma}). Then, (\ref{eq:denslim})  is consistent with a single large condensate realizing the current deviation and (\ref{eq:phi2lim}) determines the convergence of $\phi_2$ towards $\phi_c$ with increasing $L$. 
Note also that the speed \eqref{eq:speedbis} of such profiles vanishes 
\begin{equation}
v_s=\frac{\phi_2^c-\phi_1^c}{R\left(\phi_2^c\right)-R\left(\phi_1^c\right)}\to 0\quad\mbox{as }L\to\infty ,
\end{equation}
since $R\left(\phi_2^c\right) /\phi_2^c \to\infty$, which is consistent with a condensed state.
In this case, for bounded jump rates with diverging density $R(\phi )$, the condensed profile can be realised as a formal limit of a travelling wave profiles with $\phi_2\to\phi_c$.  This provides a connection between suboptimal travelling waves and condensed profiles, and in-particular implies that
\begin{equation}
E_{tw} (j)\leq E_c (j) \quad\mbox{for all }j_{min} \leq j\leq J(\rho )
\label{costine}
\end{equation}
and the result \eqref{mainres1} applies. 
This is illustrated for the constant rate ZRP in Figure \ref{fig:costCRTAZRP} in Section \ref{sc:CRTAZRP}, where the optimal travelling wave profile leads actually to a strictly lower cost unless we condition on a current $j=j_{min} =0$.

In case $\rho_c < \infty$ we will see in Section \ref{sc:condTAZRP} that the rate function $I_L(j)$ can be given by the lower convex hull of the condensed and travelling wave costs as in \eqref{mainres2}.


%
%
%
%
\section{Large deviation results for different models}\label{sec:lddf}

In this section, we determine the optimal travelling wave profiles for different types of jump rates introduced in Section \ref{subsec:genexemp}, finding explicit or numerical solutions to the minimization \eqref{eq:minsystem} for travelling wave profiles, which turn out to be unique in all cases as long as the conditioned current $j$ is admissible. This unique solution depends on the parameters $j$ and $\rho$, and is denoted $\left(\phi_1^{o} ,\phi_2^{o}\right)$ in the following and also referred to as the optimal pair or fugacities. 
In light of (\ref{mainres2}), we compare the resulting cost \eqref{twcost} with the condensed cost \eqref{condcost} to derive the large deviation rate function for the current $I(j)$, and also include remarks on finite size versions $I^L (j)$ where appropriate.\\

%
%

\subsection{Constant Rate TAZRP}\label{sc:CRTAZRP}

\begin{figure}[t]
\begin{center}
\mbox{\includegraphics[width=0.46\textwidth]{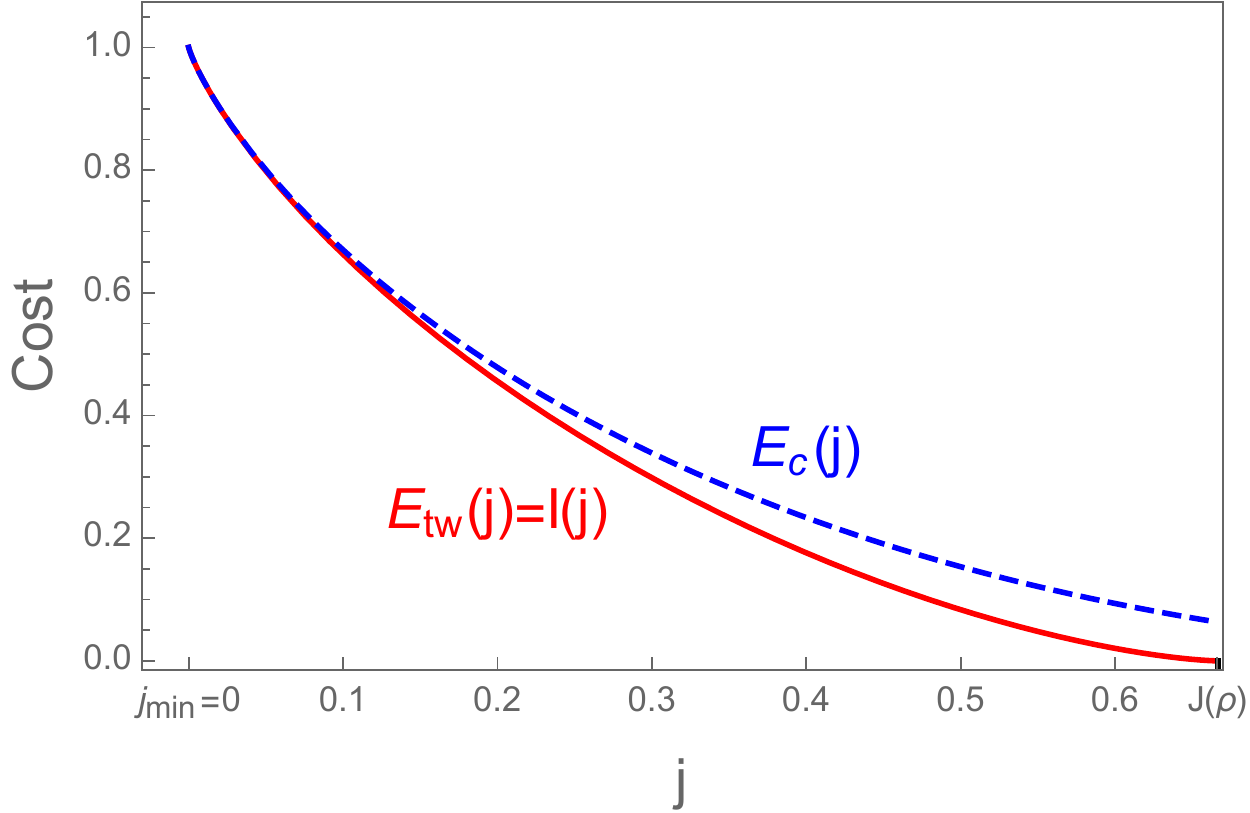}\quad\includegraphics[width=0.54\textwidth]{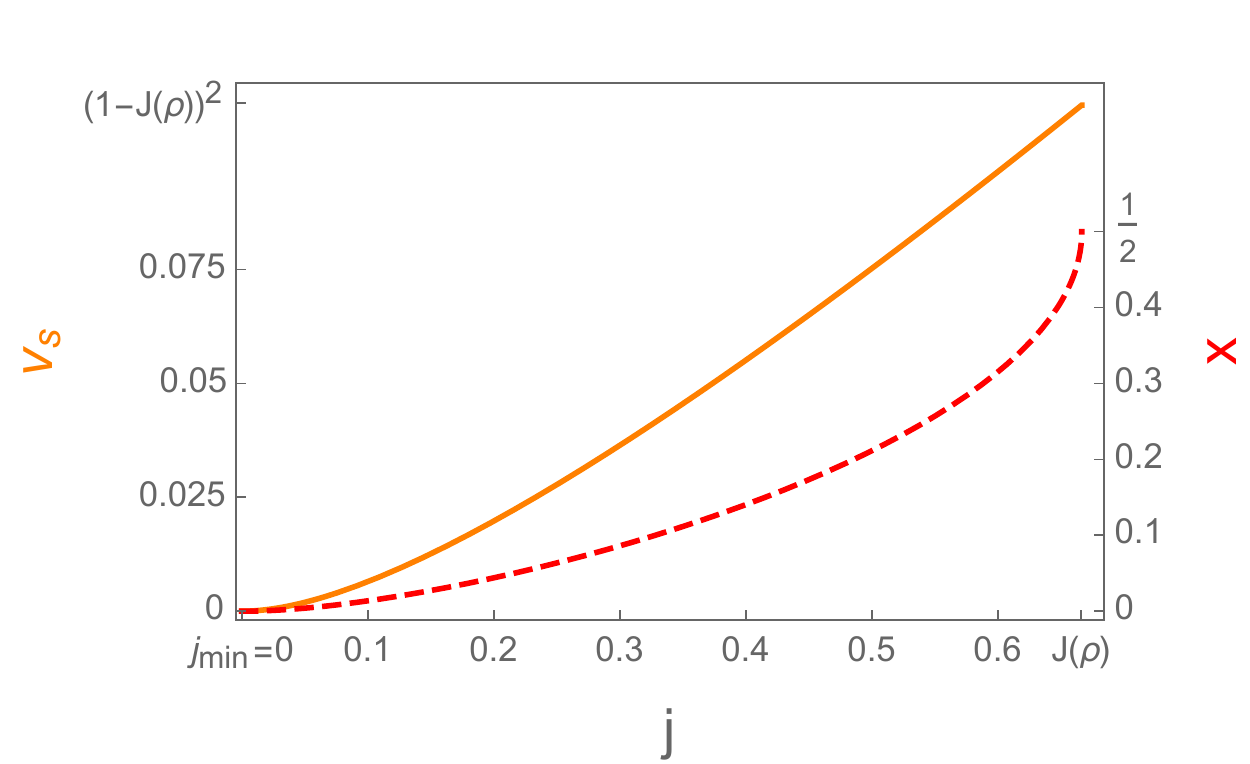}}
\end{center}
\caption{\label{fig:costCRTAZRP}
Both plots feature the constant rate TAZRP with $u\left(n\right)=1$ and $\rho=2$. (Left) The traveling wave cost $E_{tw}$ \eqref{twcost} shown in full red and the condensed cost $E_c$ \eqref{crc} in dashed blue. The condensed cost is always larger than the traveling wave cost for bounded rates. The red curve was generated implicitly from \eqref{eq:minCR} and \eqref{eq:consCR}. 
(Right) The spatial fraction of the high density phase (red dashed) and the shock speed (full orange) are increasing functions of $j$. 
At the typical current $j=J\left(\rho\right)$ we have $\phi_1^o=\phi_2^o$, the high and low density phases are indistinguishable and they occupy half of the system each, that is $x=\frac{1}{2}$. The limiting speed is given by $v_s\left(J\left(\rho\right),J\left(\rho\right)\right)=\left(1-J\left(\rho\right)\right)^2$.
}
\end{figure}

For constant rate ZRPs, with rates \eqref{rates_cr}, we have $z(\phi )=(1-\phi )^{-1}$ and $R(\phi )=\phi/(1-\phi )$ (see \eqref{eq:crrelations}), so the Jensen-Varadhan functional \eqref{jv} takes the simple form
\begin{equation}
F\left(\phi_{1},\phi_{2}\right)=\left(\phi_{2}-\phi_{1}\right)+\phi_{1}\phi_{2}\ln\frac{\phi_{1}}{\phi_{2}}-\left(1-\phi_{1}\right)\left(1-\phi_{2}\right)\ln\frac{1-\phi_{1}}{1-\phi_{2}},
\end{equation}
and the constraint (\ref{cond}) reduces to
\begin{equation}\label{eq:consCR}
G\left(\phi_1,\phi_2\right)=\phi_{1}\phi_{2}+\rho\left(\phi_{2}-1\right)\left(\phi_{1}-1\right) =j.
\end{equation}
Explicit computations of the second derivative and the determinant of the Hessian show that $\bar{\phi_2}\left(\phi_1\right)$ from (\ref{eq:consCR}) is convex and $F$ has concave level lines, which leads to unique optimal pairs $\left(\phi_1^{o} ,\phi_2^{o}\right)$. Using the above explicit expressions, the first equation in the system (\ref{eq:minsystem}) can be simplified to the implicit relation
\begin{equation}\label{eq:morsol}
\left(\phi_{2}^{o}\right)^{\rho}\left(1-\phi_{2}^{o}\right)=\left(\phi_{1}^{o}\right)^{\rho}\left(1-\phi_{1}^{o}\right).
\end{equation}
By regularity of the function $f\left(s\right)\coloneqq s^{\rho}\left(1-s\right)$, it is easy to show that \eqref{eq:morsol} has exactly one solution $\phi_2^{o} >\phi_1^o\in\left(0,1\right)$. In \cite{DerriBodJV} Section VII, a particular parametrization is given as 
\begin{equation}
\begin{array}{cc}
\phi_{1}^{o}=\frac{e^{\lambda}-e^{\lambda\left(1-\hat{\rho}\right)}}{e^{\lambda}-1} \ ,\quad \phi_{2}^{o}=\frac{e^{\lambda\hat{\rho}}-1}{e^{\lambda}-1}\ .
\end{array}
\end{equation}
Here $\lambda$ is the usual Lagrange multiplier of the maximization problem of the Jensen-Varadhan functional constrained to \eqref{rcond} and $\hat{\rho}$ is the density of the TASEP which is equivalent to the TAZRP with $\hat{\rho}=\frac{\rho}{1+\rho}$ (see Appendix \ref{sec:mapZRPEP} for a description of the mapping between the two processes). A few examples of explicit solutions to \eqref{eq:morsol} are
\begin{equation}\label{eq:minCR}
\begin{array}{cc}
\phi_{2}^{o}=\frac{1}{2}\left(2-\phi_{1}^{o}-\sqrt{\left(4-3\phi_{1}^{o}\right)\phi_{1}^{o}}\right) & \rho=\frac{1}{2}\\
\phi_{2}^{o}=1-\phi_{1}^{o} & \rho=1\\
\phi_{2}^{o}=\frac{1}{2}\left(1-\phi_{1}^{o}+\sqrt{1+2\phi_{1}^{o}-3\left(\phi_{1}^{o}\right)^{2}}\right) & \rho=2,
\end{array}.
\end{equation}
where we notice that for $\rho>1$, the $(\phi_1^o ,\phi_2^o )$ form a concave curve while for $\rho<1$ it is convex. The resulting cost function is illustrated in Figure \ref{fig:costCRTAZRP} where we plot $E_{tw}=F\left(\phi_1^o,\phi_2^o\right)$ against the current $j=G(\phi_1^o ,\phi_2^o )$.
From \eqref{eq:minCR} we see that $\phi_2^o \to 1$ as $\phi_1^o \to 0$, and in this limit $j=G\left(\phi_1^o,\phi_2^o\right)\to 0$, which is consistent with $j_{min} =0$. 
For $j\to 0$ the spatial proportion of the two phases and the shock speed are then given by
\begin{equation}\label{eq:vxCR}
x=\frac{j-\phi_{1}^{o}}{\phi_{2}^{o}-\phi_{1}^{o}}\to 0\quad\mbox{and}\quad v_{s}\left(\phi_{1}^{o},\phi_{2}^{o}\right)=\left( 1-\phi_{2}^{o}\right)\left( 1-\phi_{1}^{o}\right)\to 0\ ,
\end{equation}
as illustrated in Figure \ref{fig:costCRTAZRP}. This corresponds to a static, condensed profile, which is consistent with the weaker but more general result \eqref{limcost}, where we observe that in this case the limiting condensed profile is symptomatically optimal.
Using \eqref{eq:condcostlim} with $\phi_c =1$ the limiting cost for condensed configurations is given by
\begin{equation}
E_c (j)=1-j+j\ln j > E_{tw} (j)\quad\mbox{for all }j>0\ ,
\label{crc}
\end{equation}
and only for $j=0$ we have $E_c (0)= E_{tw} (0)=1$. 
Therefore, the large deviation rate function is given by $I (j)=E_{tw} (j)$ as shown in Figure \ref{fig:costCRTAZRP}.
%
%

\subsection{Unbounded sublinear rates}\label{sc:GUSR}

\begin{figure}[t]
\begin{center}
\mbox{\includegraphics[width=0.54\textwidth]{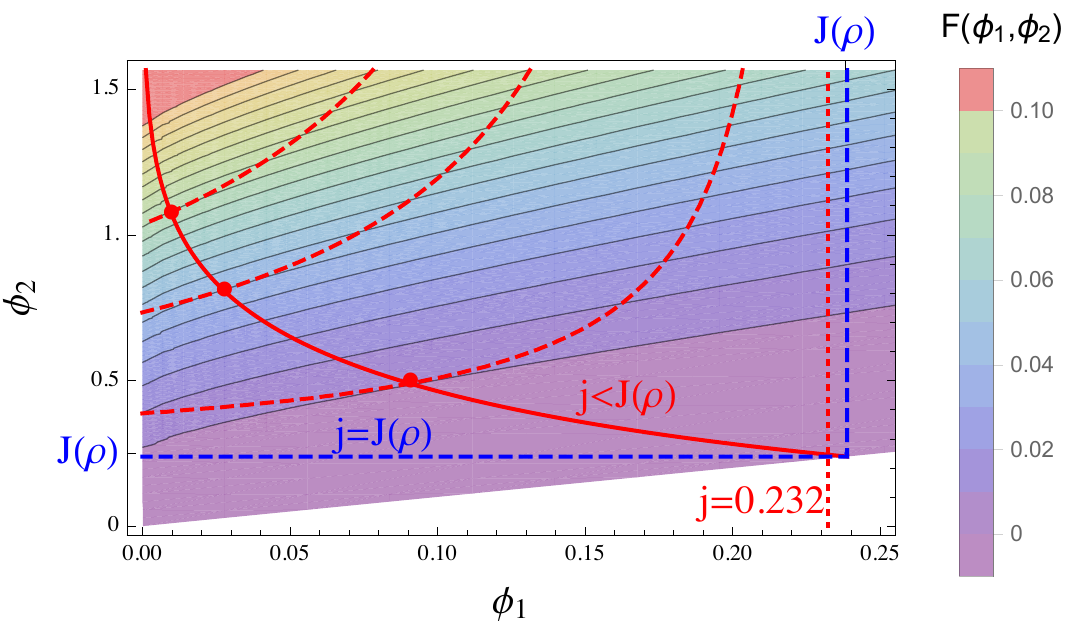}\quad\includegraphics[width=0.44\textwidth]{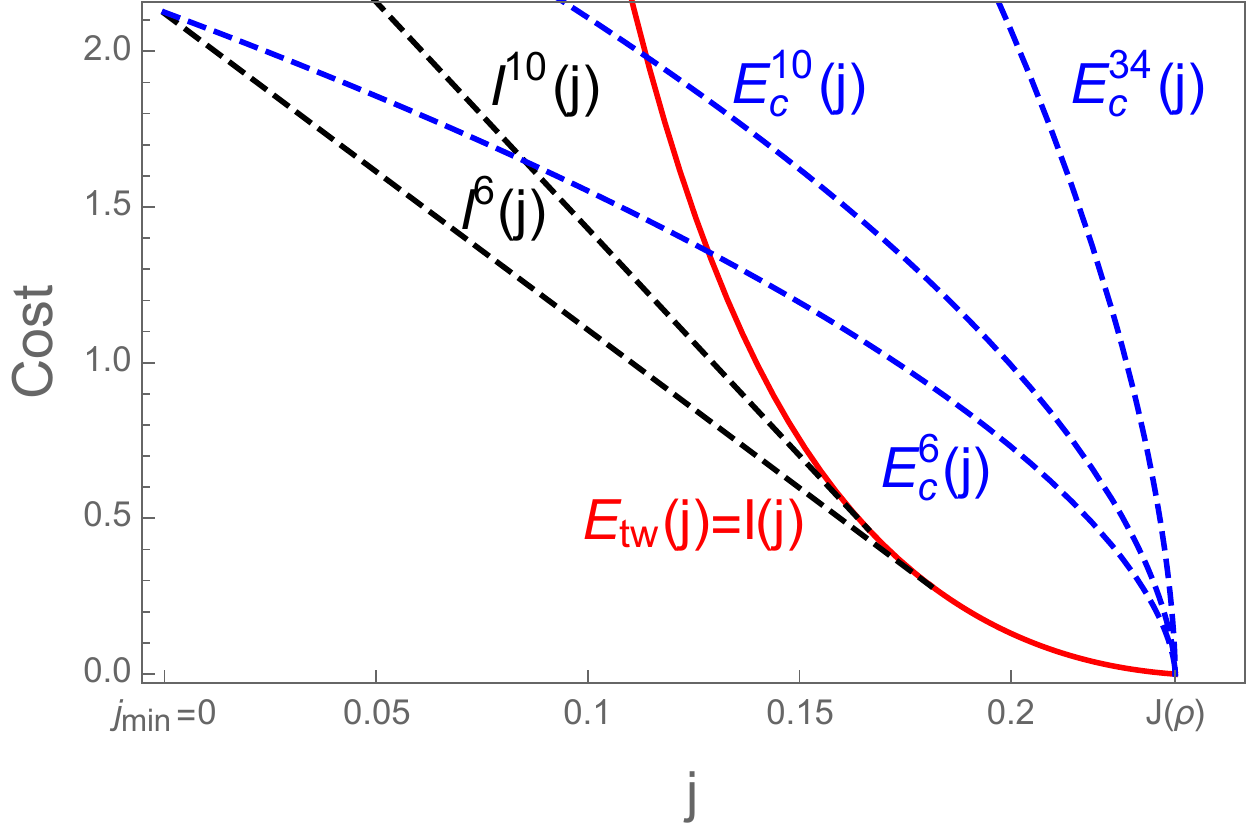}}
\end{center}
\caption{\label{fig:illu3}
Both plots feature the ZRP with rates $u\left(n\right)=\frac{(n+1)^{\gamma}-1}{\gamma}$, using $\rho=0.25$ and $\gamma=0.6$. (Left)  
Contour plot of the Jensen-Varadhan functional \eqref{jv}, constraint curves \eqref{cond} in red dashed for three values of $j<J(\rho )$. Note that all values of $j$ are close to $J(\rho )$, and the asymptote is shown for the rightmost constraint curve with $j=0.232$ (dotted red). Optimal pairs (full red) are shown analogously to Figure \ref{fig:illu}.   
(Right) The cost $E_{tw} (j)$ \eqref{twcost} (full red), diverges as $j\to j_{min}=0$, shown alongside $E^L_c (j)$ \eqref{eq:ccost} (dashed blue) for several small values of $L$. Resulting finite size rate functions $I^L (j)$ \eqref{ratefu} are approximated by dashed black lines, while the limiting rate function is equal to $I(j)=E_{tw} (j)$ in accordance with \eqref{mainres1}.
}
\end{figure}

In this section we focus on the TAZRP with rates given by $u\left(n\right)=\frac{(n+1)^{\gamma}-1}{\gamma}$ with $\gamma\in\left(0,1\right)$ introduced in \eqref{rates_sublin}, for which we have $J(\rho )\simeq (1-\rho )^\gamma /\gamma$. This implies
\begin{equation}
J(\rho )/\rho \to 0\quad\mbox{and}\quad \frac{\rho\partial_\rho J(\rho )}{J(\rho )}\to \gamma <1\quad\mbox{for}\ \rho\to\infty\ ,
\label{eq:limj}
\end{equation}
and all the results of this section will hold under these more general conditions. For the above rates, the Jensen-Varadhan functional can in general not be written as an explicit function of $\phi_1$ and $\phi_2$ and we rely on numerical solutions to calculate the optimal pairs $\left(\phi_1^{o},\phi_2^{o}\right)$ and the cost $E_{\rm tw}(j)$. 
Illustrations are shown in Figure \ref{fig:illu3} for $\gamma =0.6$. 
As $j\to j_{min}=0$ we have $\phi_1^o\to 0$ and $\phi_2^o\to\phi_c =\infty$. 
Together with \eqref{eq:hdensfrac}, this again implies that the volume fraction $x$ of the high density phase vanishes in the limit $j\to 0$ as well as the speed $v_s$ of the profile. 
Continuity of the Jensen-Varadhan functional $F$ allows us to commute limits, and formally we get
\begin{equation}\label{eq:JVunb}
\lim_{j\to 0} F\left(\phi_{1}^o ,\phi_{2}^o \right)=\lim_{\phi_{2}^o\to\infty} F(0,\phi_2^o )=
\lim_{\phi_{2}^o\to\infty}\phi_{2}^o\left(1-\frac{\ln z\left(\phi_{2}^o\right)}{R\left(\phi_{2}^o\right)}\right) =\infty\ .
\end{equation}
Here we have used l'H\^opital's rule and a change of variables to get
\begin{equation}
\lim_{\phi\to\infty}\frac{\ln z\left(\phi\right)}{R\left(\phi\right)}=\lim_{\rho\to\infty} \frac{\rho\partial_\rho J(\rho )}{J(\rho )}<1
\label{eq:}
\end{equation}
where we used $R(\phi)=\phi\partial_{\phi}\ln z(\phi)$ and the fact that $R(\phi)$ is the inverse of $J(\rho)$. The final inequality is from \eqref{eq:limj}. 

As in the previous section, for large finite systems the relevant travelling wave profiles as $j\to 0$ correspond to a high density volume fraction $x=1/L$ in \eqref{eq:hdensfrac}. This implies $R(\phi_2^o) \sim \rho L$ and a single site contains a non zero fraction of the total mass, so that $\phi_2^o \sim L^\gamma \rho^\gamma /\gamma$. Together with \eqref{eq:JVunb} this leads to a scaling of $F(0,\phi_2^o )\simeq (1-\gamma) \phi_2^o \sim L^\gamma$.

The cost for condensed profiles for large $L$ is approximately given by \eqref{condcost}, which implies
\begin{equation}
E_{c}^{L}\left(j\right)\approx\frac{1}{\gamma}\left(\left(\rho-R\left(j\right)\right)L\right)^{\gamma} =\frac{L^\gamma}{\gamma} \left(\rho-R\left(j\right)\right)^\gamma
\label{eq:ccost}
\end{equation}
for all $j<J(\rho )$. 
This is also proportional to $L^\gamma$, and again travelling wave profiles are asymptotically similar to condensed profiles with a cost on the same scale as $j\to 0$. 

As can be seen from Figure \ref{fig:illu3}, the cost for condensed profiles for all fixed $j>0$ is again higher than the one for travelling wave profiles for large enough system size. Therefore the limiting rate function is simply $I(j)=E_{tw} (j)$ and \eqref{mainres1} holds. For finite systems with fixed large $L$, however, the condensed cost $E_c^L (j)$ is eventually lower than $E_{tw} (j)$ for small enough $j$, and is a concave function of $j$. 
This leads to a linear part of the rate function $I^L (j)$ for small $j$ indicating a mixture between travelling wave and completely condensed profiles where all particles are trapped on a single site. This feature is a rather persistent finite size effect illustrated by dashed lines in Figure \ref{fig:illu3} (right). Note that the very small systems shown in the plot only contain of the order of $1$ or $2$ particles and are just intended for illustration. Low enough deviations in larger systems are not accessible numerically, so the crossover is hard to observe in simulations. 


%
%
\subsection{Asymptotically linear rates}\label{sc:ALR}
\begin{figure}[t!]
\begin{center}
\mbox{\includegraphics[width=0.48\textwidth]{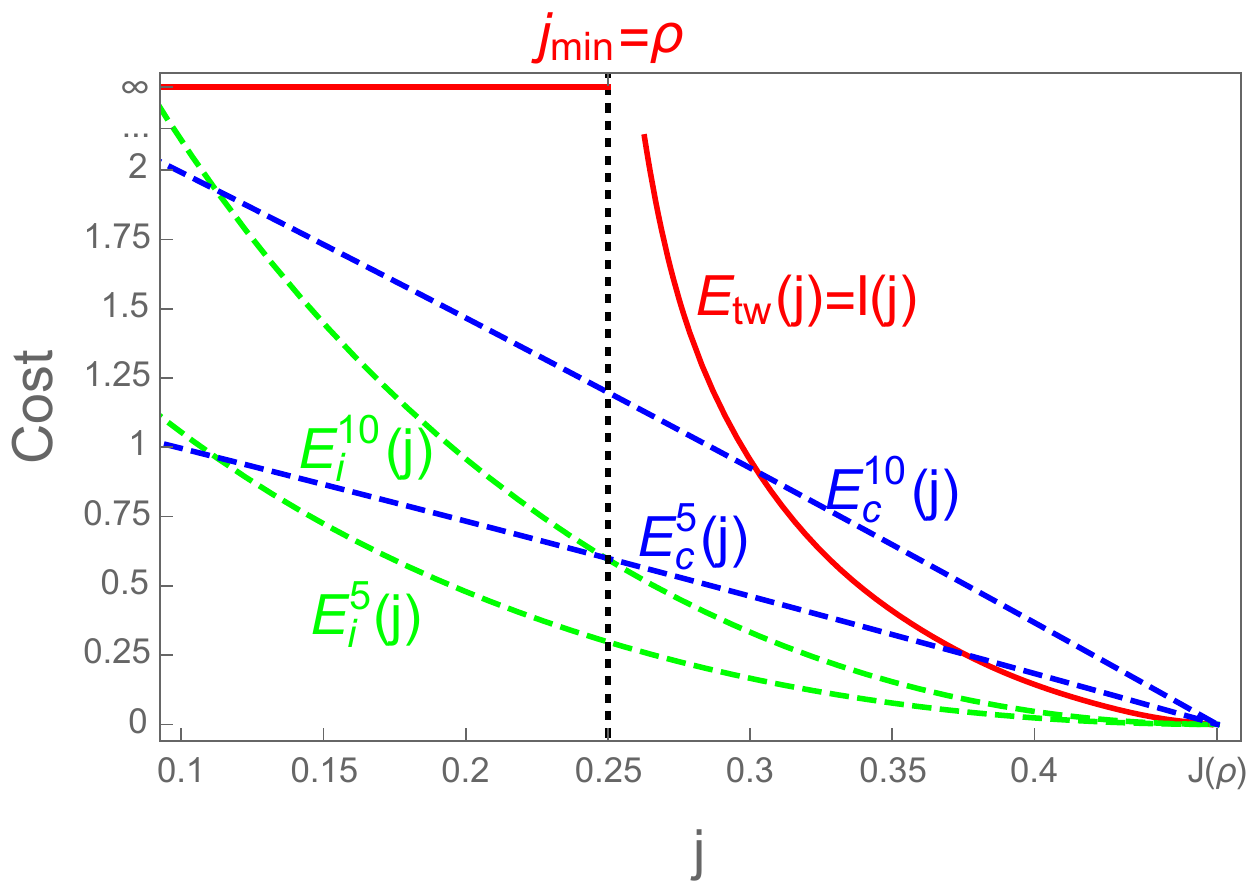}\quad\includegraphics[width=0.48\textwidth]{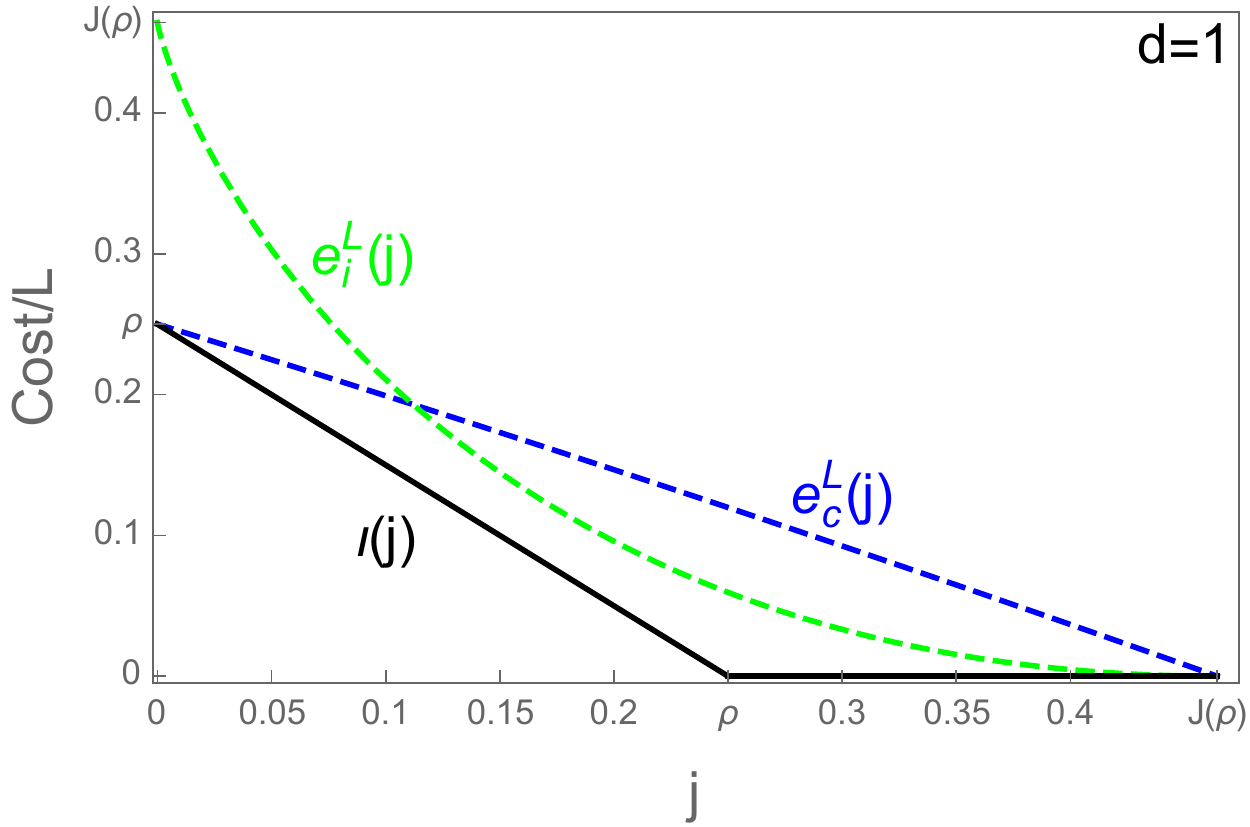}}\\
\mbox{\includegraphics[width=0.48\textwidth]{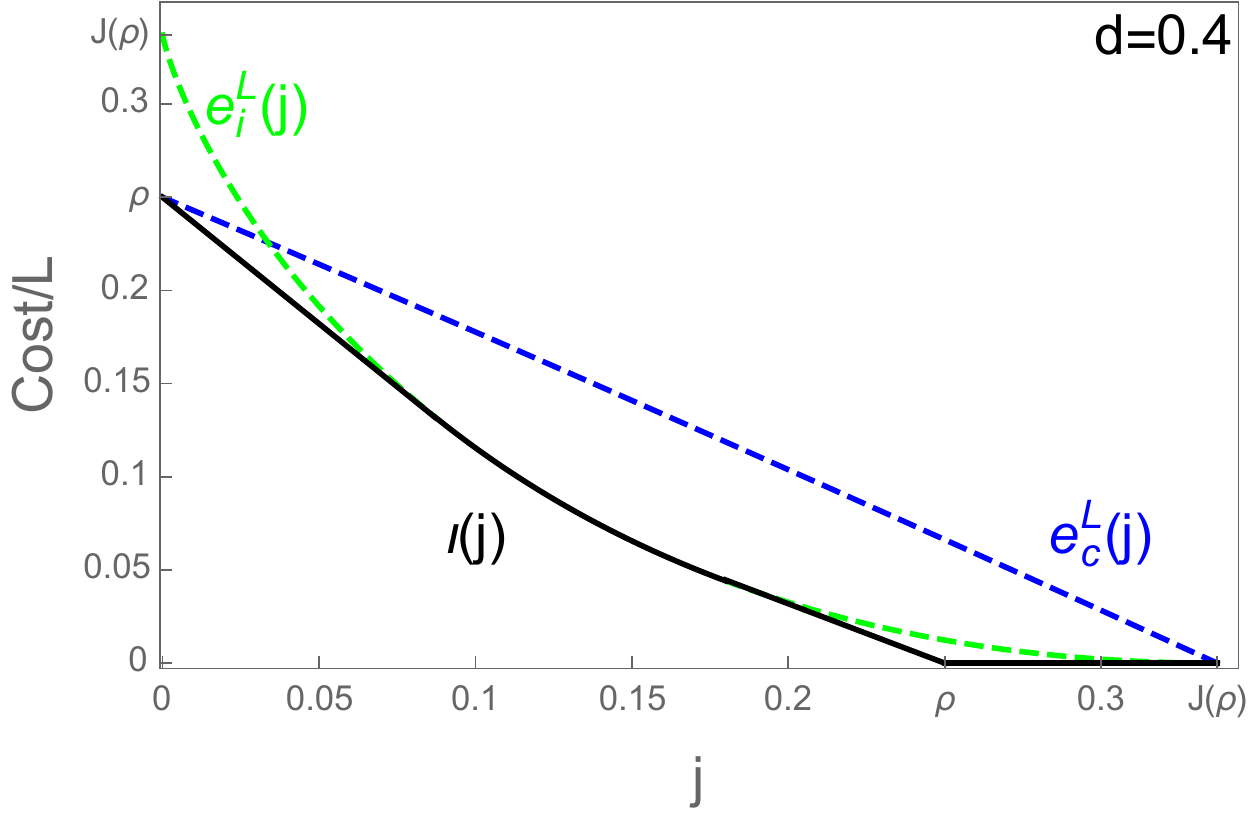}\quad\includegraphics[width=0.48\textwidth]{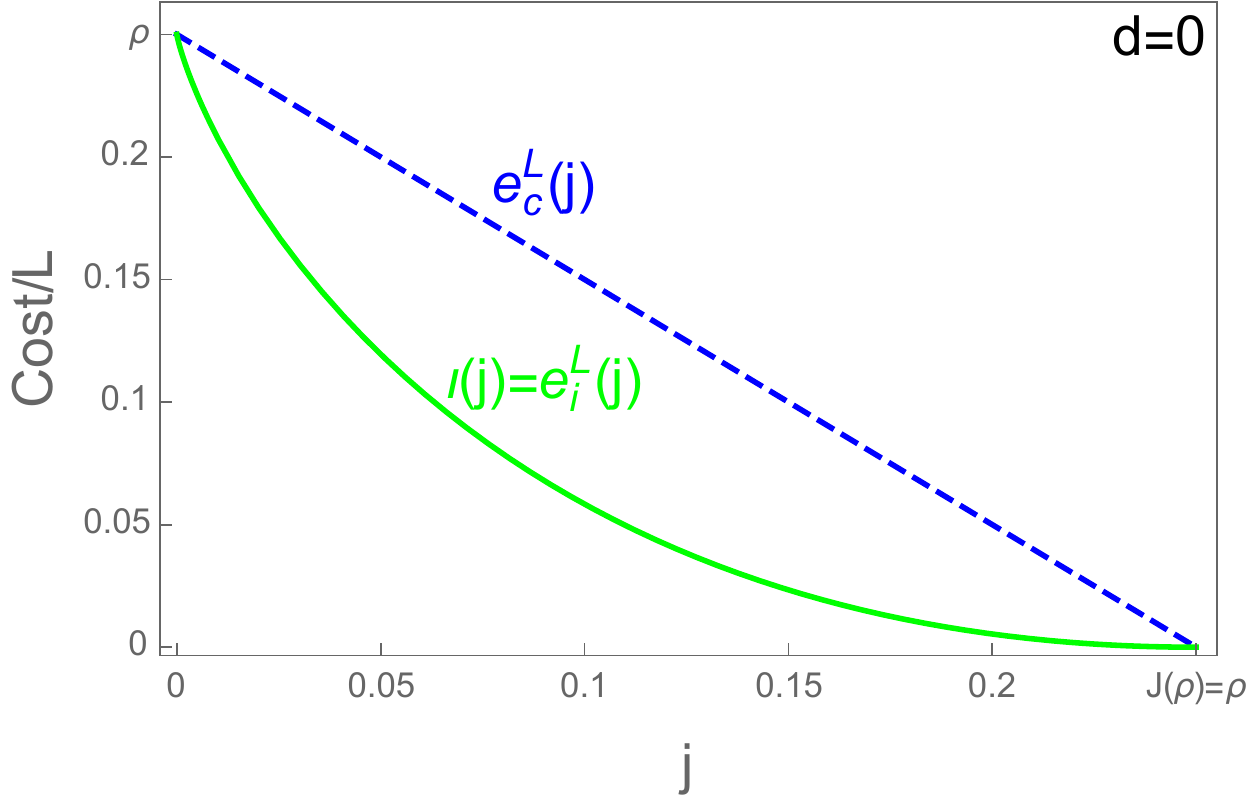}}
\end{center}
\caption{\label{fig:illu4}
%
All plots feature the ZRP with rates  $u\left(n\right)=n+d$ and $\rho =0.25$. (Top left) 
The cost is plotted against the current for $d=1$, and in accordance with \eqref{mainres1} the rate function is given by $I(j)=E_{tw} (j)$ (full red). The costs $E^L_c (j)$ \eqref{condcoin} and $E^L_i (j)$ \eqref{indi} are shown for small $L$ in dashed blue and green lines, respectively. The other plots illustrate the modified LDP \eqref{eq:modldp} with speed $Lt$ for different values of $d\geq 0$, showing the rescaled costs $e_c (j)$ (blue) and $e_i (j)$ (green), and the resulting limiting rate function $\iota (j)$ as a full black line for $d>0$. For independent particles with $d=0$ (bottom right), the rate function is dominated completely by $e_i (j)$ (green).
}
\end{figure}

Consider $u\left(n\right)=n+d$ as introduced in \eqref{rates_ind}, where $R(\phi )/\phi \to 1$ as $\phi\to\phi_c=\infty$ and with \eqref{eq:jmin} we have $j_{min}=\rho$. 
As an example in Figure \ref{fig:illu4} we consider $d=1$, and using \eqref{zphigamma} 
in this case we have the following explicit expressions 
\begin{equation}
\begin{array}{cc}
\ln z\left(\phi\right)=\ln\frac{e^{\phi}-1}{\phi} & R\left(\phi\right)=\phi-1+\frac{\phi}{e^{\phi}-1}\quad \textrm{for}\ \ d =1 \ .\end{array}
\end{equation}
As in (\ref{eq:JVunb}), the travelling wave cost diverges in the limit of $\phi_2^o \to\phi_c =\infty$. 
Furthermore, we have that $x\to 0$ and $v_s\to 1$ as $j\to j_{min} =\rho$, so in this case the travelling wave profiles in the limit $j\to j_{min}$ do not correspond to a condensed profile with a spatially fixed condensate. We do not show a contour plot of the Jensen-Varadhan functional \eqref{jv}, since it looks qualitatively the same as the one in Figure \ref{fig:illu3} for general unbounded rates, with the exception that constraint curves \eqref{cond} are defined only for $\phi_1 <j-\rho$ and exist up to currents $j\geq j_{\min} =\rho$.

Using \eqref{condcost} the condensed cost $E_c^L (j)$ increases linearly in the system size for large $L$ as
\begin{equation}\label{condcoin}
E_{c}^L\left(j\right) =u\big( (\rho -R(j))L\big)
\approx (\rho-R(j))L\ .
\end{equation}
So as long as $j>\rho$ phase separated states with an $L$-independent cost dominate the rate function and we have
\begin{equation}
I(j)=\left\{\begin{array}{cl} E_{tw} (j) &,\ j\in (\rho ,J(\rho )]\\ \infty &,\ j\in [0, \rho ]\end{array}\right.\ ,
\label{rflin}
\end{equation}
in accordance with \eqref{mainres1}. 
As in the previous section, on finite systems we expect condensed profiles to also be relevant for small currents. For this system in fact a modified large deviation principle with speed $Lt$ instead of $t$ holds in the limit $L\to\infty$, which is illustrated in Figure \ref{fig:illu4} together with \eqref{rflin}.

Since the condensed cost is of order $L$ we also have to compare to the option of slowing down the jump rate at all lattice sites which is always of order $L$ and therefore irrelevant in other examples. 
This cost is approximately given by
\begin{equation}
E_i^L (j)=L\Big( J(\rho ) -j+j\ln\frac{j}{J(\rho )}\Big)\ ,
\label{indi}
\end{equation}
where we simply approximate the integrated current out of each site by a Poisson process with rate $J(\rho )$. This is equivalent to slowing down the clock of the entire process. 
Comparing with the cost for condensed profiles, it turns out that $E_i^L (j)<E_c^L (j)$ for a range of $j$ large enough (depending on the parameter $d$), and as $j\to 0$ we have $E_c^L (0)=L\rho <LJ(\rho )=E_i^L (0)$. This is illustrated in Figure \ref{fig:illu4} (bottom row) for two parameter values $d>0$. This crossover enters the rate function of the modified LDP with speed $tL$. In this scaling, the cost of travelling wave profiles is
\begin{equation}
E_{tw} (j)/L\to e_{tw} (j):=\left\{\begin{array}{cl} 0 &,\ j\in (\rho ,J(\rho )]\\ \infty &,\ j\in [0, \rho ]\end{array}\right.\quad\mbox{as }L\to\infty\ ,
\label{eq:twscale}
\end{equation}
which again dominates the rate function for currents $j>\rho$. Therefore the rate function is given by the lower convex hull of 
\begin{equation}
I^L (j)/L\to \iota (j)\coloneqq \underline{\mbox{conv}}\{ e_{tw} (j),e_c (j),e_i (j)\} \quad\mbox{as }L\to\infty\ ,
\label{eq:modldp}
\end{equation}
which is illustrated by full black lines in Figure \ref{fig:illu4} (bottom row). Here
\begin{equation}
e_c (j)\coloneqq E^L_c (j)/L\quad\mbox{and}\quad e_i (j) \coloneqq E^L_i (j)/L
\label{eq:andrea}
\end{equation}
are $L$-independent expressions given in \eqref{condcoin} and \eqref{indi}. For $d$ large enough the rate function is simply linear between $j=0$ and $j=\rho$ and independent of $e_i (j)$, whereas $e_i (j)$ dominates an increasing part of the convex hull for decreasing $d$. For the degenerate limiting case of independent particles with $d=0$ we have $J(\rho )=\rho$ and therefore $e_{tw} (j)=\infty$ for all $j<J(\rho )$ and it does not contribute to the rate function. Then \eqref{eq:modldp} is given by the cost $e_i (j)$ of slowing down the clock of the process on all sites, or equivalently slowing down all independent particles as is expected in this case (see Figure \ref{fig:illu4} top right).

It is currently out of reach to numerically confirm the extensive behaviour of the rate function for $j\leq j_{min}$ for $d>0$ in reasonably large systems, but our heuristics is consistent with the case of independent particles with $d=0$, for which the rate function is exact. The cases in Figure \ref{fig:illu4} (top left) for very small system sizes are numerically accessibe but contain only between $1$ and $3$ particles, and are only shown for illustration. We do not expect the rate function measured in such systems to coincide with the lower convex hull of the costs since our theoretical arguments only apply for large enough $L$.

\subsection{Condensing TAZRP}\label{sc:condTAZRP}

\begin{figure}[t]
\begin{center}
\mbox{\includegraphics[width=0.51\textwidth]{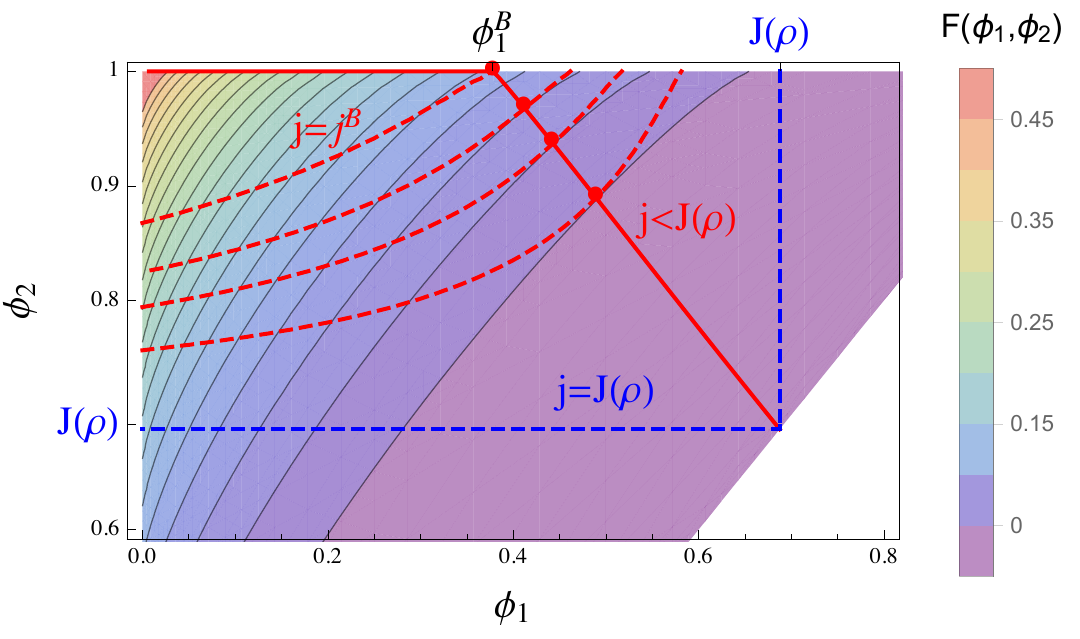}\quad\includegraphics[width=0.48\textwidth]{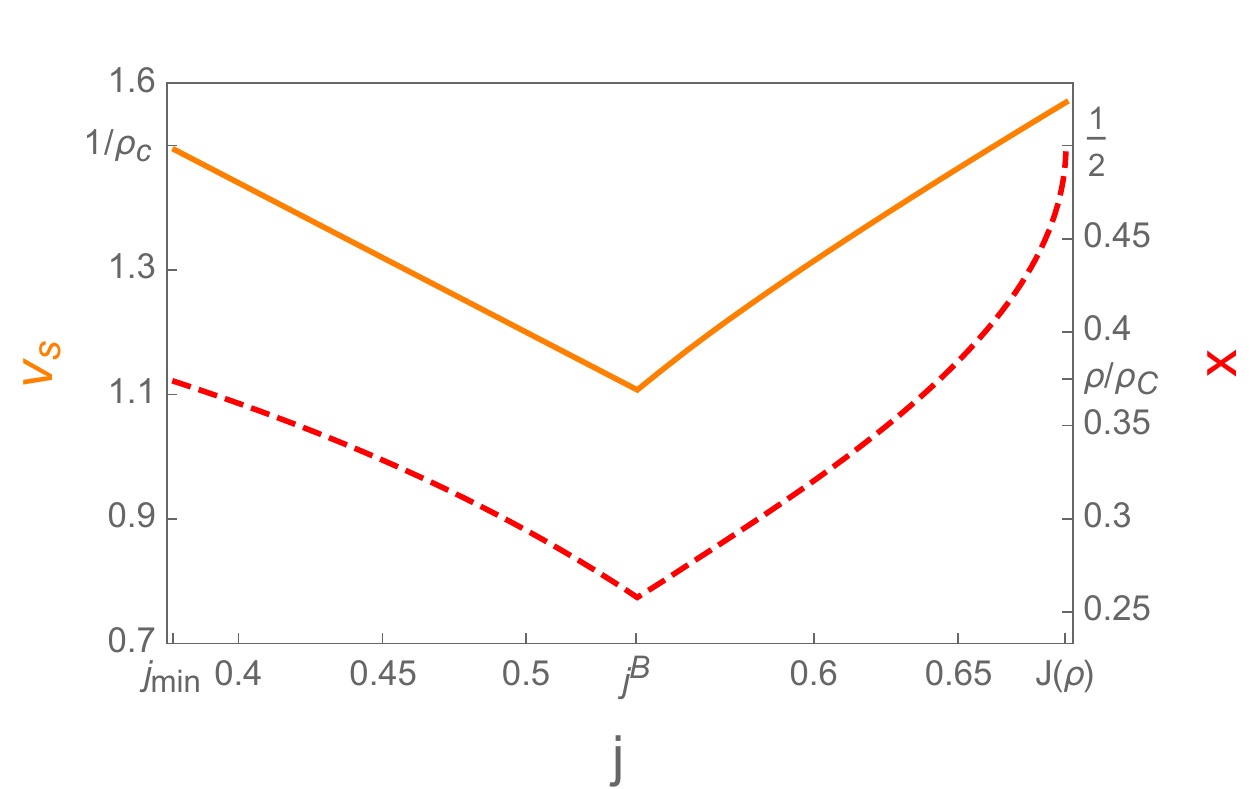}
}
\end{center}
\caption{\label{fig:illu2}
Both plots feature the ZRP with rates $u(n)=1+b/n$ \eqref{rates_cond} and parameters $b=3.5$ and $\rho=0.25$. (Left) 
The contour plot of the Jensen-Varadhan functional \eqref{jv} is shown together with the constraint curves \eqref{cond} for several values of $j < J(\rho )$ (dashed red) and optimal pairs (full red) analogous to Figure \ref{fig:illu}. For $j<j^B$ optimal pairs correspond to boundary minimizers with $\phi_2^o =1$ as explained in the text, with corresponding value $\phi_1^B < j_B$. 
(Right) The red dashed line is the optimal high density fraction $x$ \eqref{eq:hdensfrac} as a function of the conditioned current $j$, while the full orange line is the profile speed $v_s$. Note that both quantities are plotted on different scales with two ordinate axes. They are not monotone and have a minimum at $j^B$, as opposed to the constant rate case shown in Figure \ref{fig:costCRTAZRP} right.
}
\end{figure}

In this section we discuss rates $u(n)=1+b/n$ with $b>2$ as given in \eqref{rates_cond}, which exhibit condensation and have a bounded range of currents $\phi \in [0,1]$ as well as densities with $R(1)=\rho_c =1/(b-2)$. We focus on total densities $\rho <\rho_c$. The contour plot shown in Figure \ref{fig:illu2} (left) for $b=3.5$ and $\rho =0.25$ now includes the upper boundary $\phi_2 =1$ for the possible values of optimal pairs, as opposed to Figure \ref{fig:illu} for the constant rate case. The red line indicates the optimal pairs $(\phi_1^o ,\phi_2^o)$ conditioned on $j_{min} <j<J\left(\rho\right)$, where with \eqref{eq:jmin} and \eqref{rhoc} $j_{min}=\frac{\rho}{\rho_c} =\rho (b-2)<1$. For the parameters in Figure \ref{fig:illu2} there exists a current value $j^B \in (j_{min} ,J(\rho ))$ where the optimum of the Jensen Varadhan functional switches between a bulk local and a boundary minimizer with $\phi_2^o =1$. This leads to a non-monotone behaviour of the high density fraction $x$ and the speed $v_s$ of the profile, as shown in Figure \ref{fig:illu2} (right). It also leads to a kink in the cost curve $E_{tw} (j)$ at $j=j^B$. This kink is hard to observe numerically for interesting parameter values and not of particular interest as $E_{tw} (j)$ remains a convex function.

\begin{figure}[t]
\begin{center}
\mbox{\includegraphics[width=0.48\textwidth]{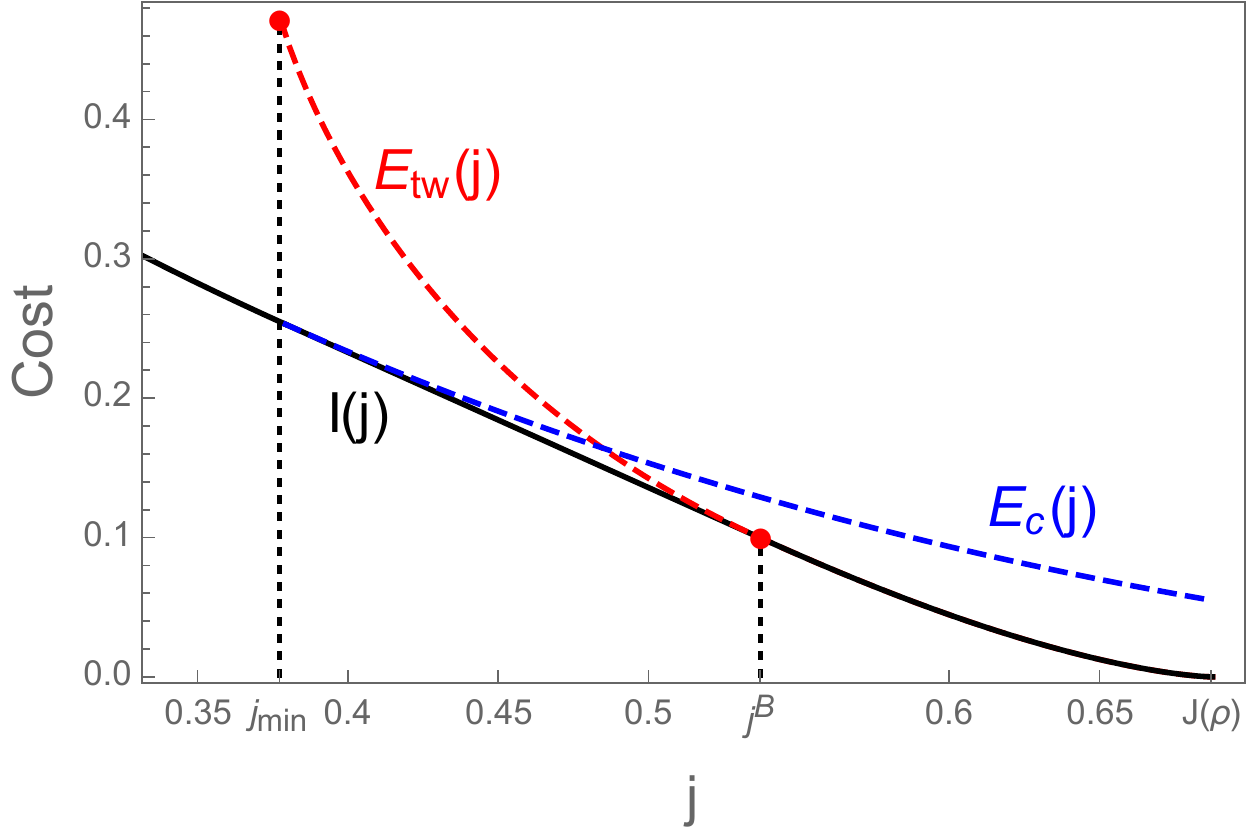}\quad\includegraphics[width=0.48\textwidth]{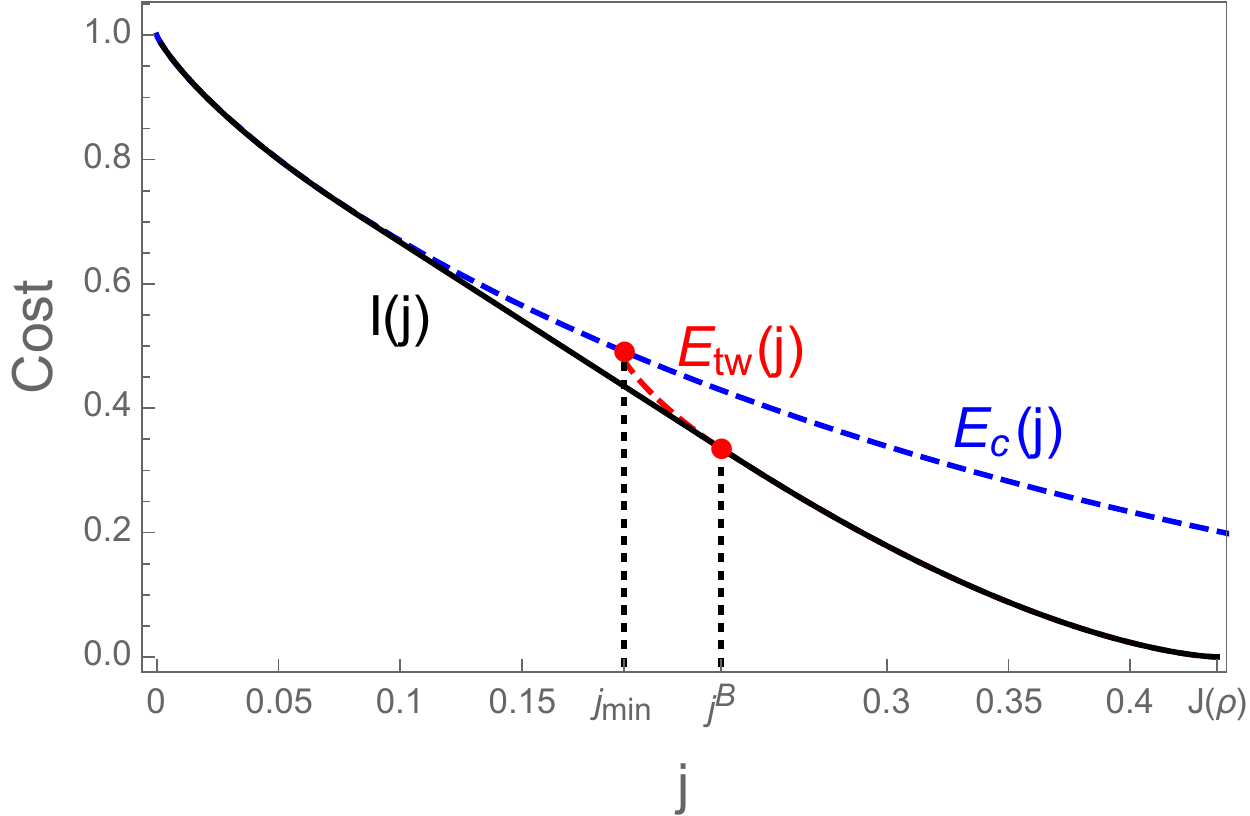}}
\end{center}
\caption{\label{fig:illu2bis} Cost functions $E_{tw} (j)$ \eqref{twcost} for travelling waves (red dashed) and $E_c (j)$ \eqref{crc} for condensed profiles (blue dashed) for rates $u\left(n\right)=1+\frac{b}{n}$ with $b=3.5$. The rate function $I(j)$ is given by the lower convex hull (full black) in accordance with \eqref{mainres2}. Note that different ranges on the axes are used. 
For $\rho=0.25$ (left) travelling wave and condensed cost curves intersect. 
For $\rho=0.12634$ (right) obtained from condition \eqref{eq:rhocondmin} $E_{tw} (j)$ and $E_c (j)$ just touch, and for smaller density values they do not intersect.
}
\end{figure}

In general, since $\phi_1^o\to 0$ as $j\to j_{min}$, the profile speed \eqref{eq:speedbis} satisfies
\begin{equation}
v_{s}=\frac{1-\phi_{1}^{o}\left(j\right)}{\rho_{c}-R\left(\phi_{1}^{o}\left(j\right)\right)}\to\frac{1}{\rho_{c}}\quad\text{as }j\to j_{min}\ ,
\end{equation}
and
\begin{equation}
x=\frac{j-\phi_{1}^{o}\left(j\right)}{1-\phi_{1}^{o}\left(j\right)}\to j_{min}=\frac{\rho}{\rho_{c}}\quad\text{as }j\to j_{min} \ .
\end{equation}
We can also again commute limits due to continuity of $F$ and get from \eqref{jv}
\begin{equation}\label{eq:limiJVcond}
\lim_{j\to j_{min}}F\left(\phi_{1}^{o}\left(j\right),\phi_{2}^{o}\left(j\right)\right)
 =F(0,1)=1-\frac{\ln z\left(1\right)}{\rho_c}=1-(b-2)\ln\frac{b}{b-1}\ ,
\end{equation}
which is finite and depends only on the parameter $b$. This is the maximum of the cost curve $E_{tw} (j)$ attained at $j=j_{min}=\rho (b-2)$ shown in Figure \ref{fig:illu2bis} for two different values of $\rho$. As in the constant rate case \eqref{crc}, the limiting condensed cost is given by the simple expression $E_c (j)=1-j+j\ln j<\infty$ independently of all system parameters and valid for all $j\in [0,J(\rho )]$. Depending on the parameters $b>2$ and $\rho <\rho_c$, the costs $E_{tw} (j)$ and $E_c (j)$ may or may not intersect, as is illustrated in Figure \ref{fig:illu2bis}. In fact, for any fixed $b>2$, there exists $\rho$ small enough such that $E_{tw}\left(j\right)\leqslant E_c\left(j\right)$ for all $j \in \left[j_{min},J\left(\rho\right)\right]$. To obtain the largest such $\rho$, we can compare \eqref{eq:limiJVcond} with the condensed cost at $j=j_{min}$ 
to obtain the condition
\begin{equation}\label{eq:rhocondmin}
\rho-\ln z\left(1\right)\leqslant\rho\ln\left(\frac{\rho}{\rho_c}\right) ,
\end{equation}
which can be solved numerically and is used in Figure \ref{fig:illu2bis} (Right).

Since $j_{min} > 0$ and the traveling wave and condensed cost both occur on the same scale, in this case the rate function is given by the non-trivial convex combination of both costs as in \eqref{mainres2}, illustrated by full black lines in Figure \ref{fig:illu2bis}. In the example plotted the right endpoint of the convex hull coincides with $j=j^B$, where $E_{tw} (j)$ exhibits a (hardly visible) kink. While the kink facilitates this behaviour, it does not hold in general and there are parameter values where the convex hull starts above or below $j^B$. 
The crossover from travelling wave profiles to condensed states in the realization of current large deviations corresponds to a dynamical phase transition. For currents $j$ in the affine region of the rate function $J(\rho )$, the large deviation is realized by a temporal mixture between travelling wave and condensed profiles in analogy to classical phase separation phenomena (see e.g.\ \cite{Touchette2009a,Touchette2014}). 
The dynamical phase transition is confirmed by numerical results presented in the next subsection, which require a detailed consideration of finite size corrections to the above arguments.

%
%
%
%
\subsection{Numerical Results for the condensing TAZRP}\label{sec:numres}

\begin{figure}[t]
\begin{center}
\mbox{\includegraphics[width=0.48\textwidth]{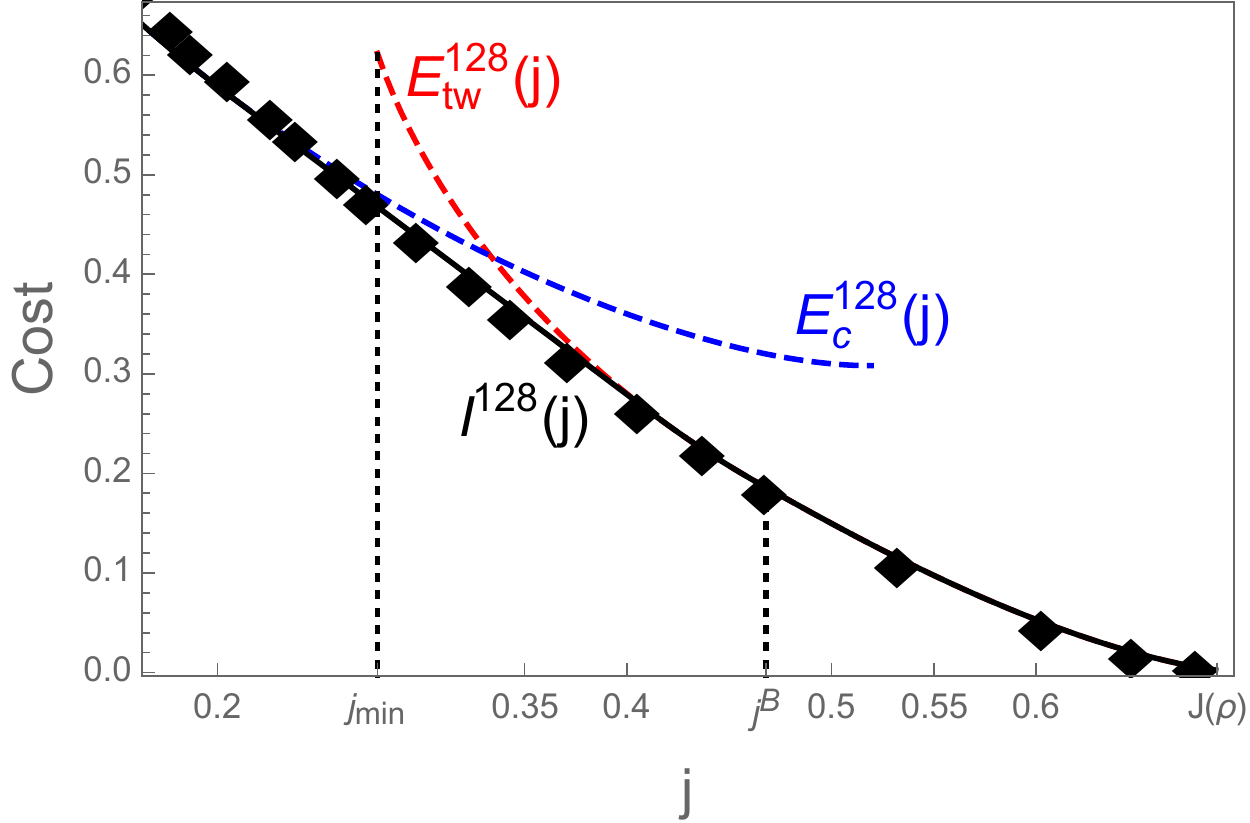}\quad\includegraphics[width=0.48\textwidth]{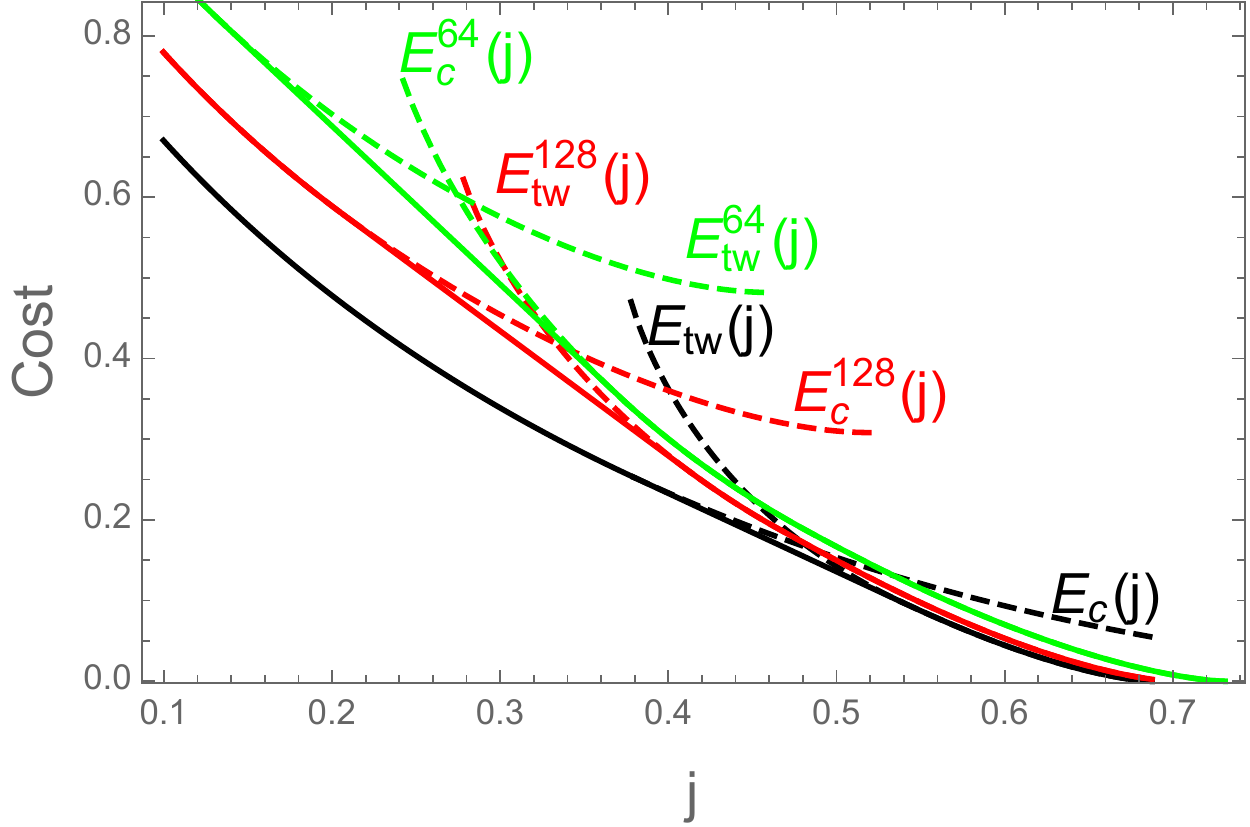}}
\end{center}
\caption{\label{fig:128cond}
Both plots feature the ZRP with rates $u\left(n\right)=1+\frac{b}{n}$ defined in \eqref{rates_cond} with parameters $\rho=0.25$ and $b=3.5$. (Left) Numerical data (black diamonds) obtained from the cloning algorithm (up to time $L^2$ with $2^{15}$ clones) are compared to finite-size cost functions $E_{tw}^L$ (red dashed) and $E_c^L$ (blue dashed) for $L=128$, and coincide very well with the predicted rate function $I^L$ given by the lower convex hull (full black). (Right) Finite-size corrections for cost functions $E_{tw}^L$ and $E_c^L$ (dashed) and the corresponding rate function $I^L$ (full) for $L=128$ (red) and $L=64$ (green) are compared with the limiting prediction (black). 
}
\end{figure}

We numerically approximate the scaled cumulant generating function $\lambda(k)$ given in \eqref{eq:momgen} using a cloning algorithm approach (see e.g.\ \cite{Giardina2011}), which is explained in Appendix \ref{sec:cloning}. 
The finite-size rate function $I^L$ is then approximated by numerically performing the Legendre-Fenchel transform \eqref{eq:leg} of the generated data. 
The results for the ZRP with rates \eqref{rates_cond} with $b=3.5$ and density $\rho = 0.25$, are shown in Figure \ref{fig:128cond} (left), and agree well with our theoretical prediction after finite size corrections. The finite-size cost functions $E_c^L (j)$ and $E_{\rm tw}^L(j)$ are defined using the canonical current density relation $J_{L,N} = \langle u \rangle_{L,N}$ with $N=[\rho L]$ as given in \eqref{cancurr}, in place of the limiting current $J(\rho)$. 
It is well known that $J_{L,N} =Z_{L,N-1}/Z_{L,N}$, and it can be computed exactly using the recursion $Z_{L,N} =\sum_{k=0}^N w(k)\, Z_{L,N-k}$ for the partition function (see e.g.\ \cite{CGfinite} and references therein). For finite $L$, the maximum current is larger than the limiting value, $\phi_c^L >\phi_c =1$, and the current is known to significantly differ from its limiting behaviour above the critical density \cite{CGfinite}. Inversion of this function defines the density $R^L (\phi )$ as a function of the current. This leads to a finite-size version of the Jensen-Varadhan functional \eqref{jv} $F^L (\phi_1 ,\phi_2 )$ and of the constraint function $G^L (\phi_1 ,\phi_2 )$, which are used as in \eqref{twcost} to define a finite-size version of $E_{tw}^L (j)$. The density $R^L (j)$ is also used in \eqref{condcost} to define a finite-size corrected version of $E_c^L$. The resulting finite size corrections to the predicted rate function are significant, as shown in Figure \ref{fig:128cond} (right). 

The simulations used to calculate the moment generating function $\lambda(k)$, are performed in an ensemble where the average integrated current is fixed by the conjugate parameter $k$, rather than conditioning the path distribution on a current $j$. Both parameters a conjugate, and the average current $j(k)$ for a given value of $k$ is given by $\partial_k \lambda(k)$. 
Affine regions of the rate function $I$ correspond to discontinuous derivatives of $\lambda(k)$, and cannot be explored by the cloning algorithm. On finite systems these effects are smoothed out somewhat, which leads to data points from the simulations also in the affine regions of the rate function. From simulations with a cloning ensemble it is not possible to directly observe temporal mixtures, which realize such large deviation events for the original ZRP conditioned on a current $j$ in the affine region of the rate function. 
The slight systematic error visible in Figure \ref{fig:128cond} is due to a generic sampling bias, which is caused by finite observation times leading to under-estimation of the probability for small values of $j$, and an over-estimation for values of $j$ close to $J(\rho)$.



%
%
%
%
\section{Conclusion and Outlook}

We study lower current large deviations for general TAZRP with concave flux functions $J(\rho )$, which can be realized by phase separated density profiles. Travelling wave profiles related to non-entropic hydrodynamic shocks are identified as the universal typical realization at least for small deviations from the typical current. These shocks can be stabelized by local changes in the dynamics and lead to rate functions which are independent of the system size, which have been studied before for the exclusion process. The range of accessible currents for these profiles may be limited, and we established a dynamical phase transition where large deviations for low currents are realized by condensed profiles. In this case the rate function is determined by slowing down the exit process out of the condensate which is again independent of the system size in the case of bounded rates. The transition is caused by two basic mechanisms (summarized in Figure \ref{fig:brange}); firstly, the range of densities in travelling wave profiles is bounded by the critical density in condensing ZRPs, this leads to a minimal accessibe current of $j_{min}=\rho /\rho_c$. Secondly, the ratio of limiting current and density appearing in \eqref{eq:jmin} may be bounded due to an asymptotically linear current density relation. In this case the rate function for condensed states is extensive in the system size. We have studied these cases in detail for typical examples of jump rates, together with other generic models with bounded and unbounded rates which do not exhibit a dynamic transition. In this way we cover all qualitative cases of concave flux functions which gives a complete picture of the large deviations for lower current deviations formulated in \eqref{mainres1} and \eqref{mainres2} in the limit of diverging system size. For condensing systems large deviations of the current may be realized by a temporal mixture leading to a convex rate function, which we have confirmed by numerical simulations using a cloning algorithm in Section \ref{sc:condTAZRP}. For finite systems, other strategies beyond travelling waves or condensed profiles may play a role as is illustrated for asymptotically linear rates in Section \ref{sc:ALR}.


For future works it would be desirable to complement our analysis with exact results derived from a microscopic approach, analogously to results for open boundary systems \cite{Hirschberg2015}, and to investigate how the dynamic transition can be understood in the framework of macroscopic fluctuation theory. While directly analogous results can be derived for upper large deviations when the flux function is convex, it would be interesting to see if general flux functions can at least partially be covered by our approach, or how it extends to partially asymmetric dynamics. As summarized e.g. in \cite{Chleboun2014}, more general Misanthrope processes also provide interesting candidates to study dynamic transitions for current large deviations. Condensed states may require a possibly modified structure, while travelling wave profiles depend only on the hydrodynamic behaviour of the process and are expected to apply in great generality.

%
%
%
%
\section*{Acknowledgements}
AP acknowledges support by the Engineering and Physical Sciences Research Council (EPSRC), Grant No. EP/L505110/1 and by the Italian National Group of Mathematical Physics (GNFM-INdAM) Progetto-Giovani-2016 \textit{Statistical Mechanics for Deep
Learning}. Moreover, some special thanks go to Yevgeny Vilensky for having shared his PhD thesis with us.
We are grateful for inspiring discussions with colleagues, in particular Thomas Rafferty. 

%
%
%
%

\appendix

\section*{Appendix}
\addcontentsline{toc}{section}{Appendix}

\setcounter{section}{1}

\subsection{Remarks on the cloning algorithm}\label{sec:cloning}

The cloning dynamics are motivated, and specified, as follows.
We may rewrite the moment generating function \eqref{eq:momgen} in terms of an expectation of the constant function with respect to a non probability-conserving process (see for example \cite{Lebowitz1998,Chetrite2014}). 
Precisely, for an initial configuration $\eta$ we have 
\begin{align*}
  \mathbb{E}_\eta\left[e^{tk\mathcal{J}^L}\right] = \left(e^{t\mathcal{L}_k}\bold{1}\right)(\eta)\,,
\end{align*}
where $\bold{1}$ is the constant function equal to one at every point of the state space. $\mathcal{L}_k$ (typically called the tilted generator of the process) is given by
\begin{align*}
  \mathcal{L}_kf(\eta) = \sum_{x\in\Lambda}u(\eta_x)\left[e^k f(\eta^{x,x+1})-f(\eta)\right]\,.
\end{align*}
Since the state space is finite for fixed $L$ and $N$, it follows from the Perron-Frobenius theorem that the operator $\mathcal{L}_k$, restricted to real functions on $X_{L,N}$, has a unique real eigenvalue with maximal real part, which is equal to the scaled cumulant generating function
\begin{align*}
  \lambda(k) = \lim_{t\to\infty} \frac{1}{t}\ln  \left[e^{t\mathcal{L}_k}\bold{1}(\eta_t)\right]\,.
\end{align*}
Since we focus on downward large deviations of the current, we may restrict to $k < 0$ and can rewrite the tittled generator as
\begin{align*}
  \mathcal{L}_kf(\eta) = \sum_{x\in\Lambda}u(\eta_x)\left(e^k \left[f(\eta^{x,x+1})-f(\eta)\right] + (1-e^k)\left[0-f(\eta)\right] \right)\,,
\end{align*}
which can be interpreted as a killed process: 
Particles attempt to jump off a site $x$ and move to the right neighbor at rate $u(\eta_x)$. 
With probability $e^k$ the move is successful, and with probability $(1-e^k)$ the process is killed.
Then $e^{t\mathcal{L}_k}\bold{1}(\eta)$ is given by the probability that the process started from $\eta$ has not been killed by time $t$. 

This probability can in principle be estimated by a strong law of large numbers, starting from $M_0$ independent copies of these auxiliary Markovian dynamics, $e^{t\mathcal{L}_k}\bold{1}(\eta)\approx M_t/M_0$, where $M_t$ is the number of chains which have not been killed by time $t$. 
However, $M_t$ decays exponentially quickly in $t$, so such a simulation would have to be started from an unfeasibly large $M_0$ in order to get reasonable statistics since $\lambda(k)$ is determined by the large $t$ behaviour. 
The cloning algorithm keeps a constant ensemble size $M_0$, where each time one of the chains is killed another chain is picked uniformly at random from the surviving chains, and its entire history up to the current time is copied.
Each cloning event corresponds to effectively rescaling the population by a factor $\frac{M_0}{M_0 -1}$, and hence $e^{t\mathcal{L}_k}\bold{1}(\eta) \approx \left(\frac{M_0 -1}{M_0}\right)^{C_t}$, where $C_t$ is the number of cloning events up to time $t$. 
It follows that 
\begin{align*}
  \lambda(k) \approx \frac{C_t}{t}\ln \left(\frac{M_0 -1}{M_0}\right)\,, \quad \textrm{for $t$ and $M_0$ sufficiently large.}
\end{align*}

We sample a single value of $C_t$, for large $t$, by running an (exact) Gillespie algorithm for the cloning dynamics.
The results in Figure \ref{fig:128cond} were obtained by running the dynamics up to a final time $t= L^{2}$ with $M_0 = 2^{15}$ clones.

\subsection{An auxiliary result\label{sec:lemma}}

\textbf{Lemma.} 
Consider a ZRP with critical fugacity $\phi_c\in (0,\infty ]$. 
If $R(\phi )/\phi \to\infty$ as $\phi\to\phi_c$, then
\begin{equation}\label{lemres}
\lim_{\phi\to\phi_c}\frac{\ln z\left(\phi\right)}{R\left(\phi\right)}=0.
\end{equation}

\noindent Note that for $\phi_c <\infty$ the assumption is equivalent to $R\left(\phi\right) \to \infty$ as $\phi\to\phi_c$.

\begin{proof}
Suppose for contradiction that there exists $A>0$ for which
\begin{equation}
R\left(\phi\right)\leqslant A\ln z\left(\phi\right)\quad\mbox{for all }\phi\in\left[0,\phi_{c}\right).
\end{equation}
Then, we can pick $\tilde{\phi}<\phi_c$, and $A'>0$ such that
\begin{equation}
\partial_{\phi}\ln z\left(\phi\right)\leqslant A\frac{\ln z\left(\phi\right)}{\phi}\leqslant A'\ln z\left(\phi\right)\quad\mbox{for all }\phi\in [\tilde{\phi},\phi_c ).
\end{equation}
By Gronwall's inequality this implies
\begin{equation}
\ln z\left(\phi\right)\leqslant\ln z(\tilde{\phi})\, e^{A'\left(\phi-\tilde{\phi}\right)}\quad\mbox{for all } \phi \in [\tilde{\phi},\phi_c ),
\end{equation}
and therefore
\begin{equation}
\partial_{\phi}\ln z\left(\phi\right)\leqslant A'\ln z(\tilde{\phi})\, e^{A'\left(\phi-\tilde{\phi}\right)}.
\end{equation}
This is a contradiction, since by assumption $\partial_{\phi}\ln z\left(\phi\right)\to\infty$ as $\phi\to\phi_c$.
\end{proof}

\subsection{Relation with exclusion processes\label{sec:mapZRPEP}}

Any ZRP can be mapped to an exclusion process (EP) in the following way. The number of particles $\hat{N}$ of the EP is the same as for the ZRP, that is $\hat{N}=N$, while the number of sites $\hat{L}$ of the EP is given by $\hat{L}=N+L$. Then, a site of the ZRP containing $m$ particles becomes a block of $m$ occupied sites in the EP. This is a standard mapping \cite{transcomplex}, which leads on the level of configurations to
\begin{equation}
\hat{\rho}=\frac{1}{1+\rho},
\end{equation}
where $\hat{\rho}$ is the density of particles in the EP as a function of the ZRP density $\rho$.\\ 
In this way, for any choice of the transition rates $u\left(n\right)$, the ZRP can be mapped to an EP with jump rates depending on block sizes. \\
The current per site $\hat{j}$ of the EP is simply given by a renormalization of the ZRP current per site $j$ as
\begin{equation}
\hat{j}=\frac{L}{L+N}j=\frac{\rho}{1+\rho}j.
\end{equation}
The same applies to the stationary current
\begin{equation}\label{eq:newstatcur}
\hat{J}\left(\hat{\rho}\right)=\frac{\rho}{1+\rho}J\left(\rho\right)=\left(1-\hat{\rho}\right)J\left(\frac{\hat{\rho}}{1-\hat{\rho}}\right).
\end{equation}
The EP is simply another representation of the same process. In this way, the large deviation principle \eqref{mainres2} implies the rate function
\begin{equation}
\hat{I}\left(\hat{j}\right)=I\left(j\right)=I\left(\left(1-\hat{\rho}\right)\hat{j}\right)
\label{eq:rf}
\end{equation}
for the exclusion model. As expected $\hat{I}\left(\hat{J}\left(\hat{\rho}\right)\right)=I\left(J\left(\rho\right)\right)=0$, so the rate function vanishes at the stationary current.\\
Note that, from \eqref{eq:newstatcur}, $\hat{J}\left(\hat{\rho}\right)$ is concave, since we assumed that $J\left(\rho\right)$ is concave as well \eqref{jass}. Also, for all sublinear currents $J\left(\rho\right)$, $\hat{J}\left(\hat{\rho}\right)$ is non-monotone since $\hat{J}\left(\hat{\rho}\right)=\left(1-\hat{\rho}\right)J\left(\frac{\hat{\rho}}{1-\hat{\rho}}\right)\to0$ as $\hat{\rho}\to1$. For asymptotically linear currents, like $J\left(\rho\right)\simeq d+\rho$ (see Section \ref{subsec:genexemp}), we have $\hat{J}\left(\hat{\rho}\right)\to1$ as $\hat{\rho}\to1$. Furthermore, travelling wave profiles in the ZRP map to travelling waves in the EP with shock speed $\hat{v}_s=\frac{\hat{J}\left(\hat{\rho_2}\right)-\hat{J}\left(\hat{\rho_1}\right)}{\hat{\rho_2}-\hat{\rho_1}}$. Condensed states in the ZRP also map to travelling wave profiles in the EP (which does not have condensed profiles), with the condensate corresponding to a block of fully occupied sites.\\
The concavity of $\hat{J}\left(\hat{\rho}\right)$ leads to lower current deviations being realized in the EP by phase separated profiles analogously to ZRP. \eqref{eq:rf} is then consistent with the Jensen-Varadhan approach applied directly to the exclusion representation of the system 
(as is done in \cite{DerriBodJV} for the standard TASEP, which can be mapped to the constant rate ZRP). 


%
%
%
%



\label{Bibliography}




\begin{thebibliography}{10}

\bibitem{lazar1}
A.~Lazarescu.
\newblock The physicist's companion to current fluctuations: one-dimensional
  bulk-driven lattice gases.
\newblock {\em Journal of Physics A: Mathematical and Theoretical},
  48(50):503001, 2015.

\bibitem{gorissen2012exact}
M.~Gorissen, A.~Lazarescu, K.~Mallick, and C.~Vanderzande.
\newblock Exact current statistics of the asymmetric simple exclusion process
  with open boundaries.
\newblock {\em Physical Review Letters}, 109(17):170601, 2012.

\bibitem{Bodineau2006}
T.~Bodineau and B.~Derrida.
\newblock {Current large deviations for asymmetric exclusion processes with
  open boundaries}.
\newblock {\em Journal of Statistical Physics}, 123:277--300, 2006.

\bibitem{derrida2007non}
B.~Derrida.
\newblock Non-equilibrium steady states: fluctuations and large deviations of
  the density and of the current.
\newblock {\em Journal of Statistical Mechanics: Theory and Experiment},
  2007(07):P07023, 2007.

\bibitem{Harris2005}
R.~J. Harris, A.~R{\'{a}}kos, and G.~M. Sch{\"{u}}tz.
\newblock {Current fluctuations in the zero-range process with open
  boundaries}.
\newblock {\em Journal of Statistical Mechanics: Theory and Experiment},
  2005(08):P08003--P08003, 2005.

\bibitem{Harris2013}
R.~J. Harris, V.~Popkov, and G.~M. Sch{\"u}tz.
\newblock Dynamics of instantaneous condensation in the zrp conditioned on an
  atypical current.
\newblock {\em Entropy}, 15(11):5065--5083, 2013.

\bibitem{Hirschberg2015}
O.~Hirschberg, D.~Mukamel, and G.~M. Sch{\"u}tz.
\newblock Density profiles, dynamics, and condensation in the zrp conditioned
  on an atypical current.
\newblock {\em Journal of Statistical Mechanics: Theory and Experiment},
  2015(11):P11023, 2015.

\bibitem{Lebowitz1998}
J.~L. Lebowitz and H.~Spohn.
\newblock A gallavotti-cohen type symmetry in the large deviation functional
  for stochastic dynamics.
\newblock {\em Journal of Statistical Physics}, 95(1-2):333--365, 1999.

\bibitem{harris75breakdown}
R.~J. Harris, A.~R{\'a}kos, and G.~M. Sch{\"u}tz.
\newblock Breakdown of Gallavotti--Cohen symmetry for stochastic dynamics,
  2006.
\newblock {\em Europhysics Letters}, 75:227.

\bibitem{rakos2008range}
A.~R{\'a}kos and R.~J. Harris.
\newblock On the range of validity of the fluctuation theorem for stochastic
  markovian dynamics.
\newblock {\em Journal of Statistical Mechanics: Theory and Experiment},
  2008(05):P05005, 2008.

\bibitem{popkov2010asep}
V.~Popkov, G.~M. Sch{\"u}tz, and D.~Simon.
\newblock Asep on a ring conditioned on enhanced flux.
\newblock {\em Journal of Statistical Mechanics: Theory and Experiment},
  2010(10):P10007, 2010.

\bibitem{tsobgni2016large}
N.~P. Tsobgni and H.~Touchette.
\newblock Large deviations of the current for driven periodic diffusions.
\newblock {\em Physical Review E}, 94(3-1):032101, 2016.

\bibitem{indiansZRP07}
S.~Gupta, M.~Barma, and S.~N. Majumdar.
\newblock Finite-size effects on the dynamics of the zero-range process.
\newblock {\em Physical Review E}, 76(6):060101, 2007.

\bibitem{Bertini2014a}
L.~Bertini, A.~De~Sole, D.~Gabrielli, G.~Jona-Lasinio, and C.~Landim.
\newblock Macroscopic fluctuation theory.
\newblock {\em Reviews of Modern Physics}, 87(2):593, 2015.

\bibitem{3bigsfirst}
L.~Bertini, A.~Faggionato, and D.~Gabrielli.
\newblock Large deviations of the empirical flow for continuous time markov
  chains.
\newblock {\em Annales de l'institut Henri Poincare (B) Probability and
  Statistics}, 51(3), 867-900. 10.1214/14-AIHP601, 2015.

\bibitem{3bigssecond}
L.~Bertini, A.~Faggionato, and D.~Gabrielli.
\newblock Flows, currents, and cycles for markov chains: Large deviation
  asymptotics.
\newblock {\em Stochastic Processes and their Applications}, 125(7), 2786-2819.
  10.1016/j.spa.2015.02.001, 2015.

\bibitem{jack2015hyperuniformity}
R.~L. Jack, I.~R. Thompson, and P.~Sollich.
\newblock Hyperuniformity and phase separation in biased ensembles of
  trajectories for diffusive systems.
\newblock {\em Physical Review Letters}, 114(6):060601, 2015.

\bibitem{karevski2016conformal}
D.~Karevski and G.~M.~Sch{\"u}tz.
\newblock Conformal invariance in driven diffusive systems at high currents.
\newblock {\em arXiv preprint}, 1606.04248, 2016.

\bibitem{Smoller}
J.~Smoller.
\newblock {\em {Shock Waves and Reaction-Diffusion Equations}}.
\newblock Springer-Verlag, 1994.

\bibitem{Varadhan2004}
S.~R.~S. Varadhan.
\newblock {Large deviations for the asymmetric simple exclusion process}.
\newblock {\em Advanced Studies in Pure Mathematics}, pages 1--27, 2004.

\bibitem{Jensen}
L.~H. Jensen.
\newblock {\em {Large Deviations of the Asymmetric Simple Exclusion Process in
  One Dimension}}.
\newblock PhD thesis, 2000.

\bibitem{Vilensky2008}
Y.~Vilensky.
\newblock {\em {Large Deviation Lower Bounds for the Totally Asymmetric Simple
  Exclusion Process}}.
\newblock PhD thesis, 2008.

\bibitem{DerriBodJV}
B.~Derrida and T.~Bodineau.
\newblock Distribution of current in nonequilibrium diffusive systems and phase
  transitions.
\newblock {\em Physical Review E}, 72, 066110, 2005.

\bibitem{spitzer70}
F.~Spitzer.
\newblock {Interaction of Markov processes}.
\newblock {\em Advances in Mathematics}, 5:246--290, 1970.

\bibitem{Andjel1982}
E.~D. Andjel.
\newblock {Invariant Measures for the Zero Range Process}.
\newblock {\em Annals of Probability}, 10(3):525--547, 1982.

\bibitem{drouffe98}
J.-M. Drouffe, C.~Godr\`{e}che, and F.~Camia.
\newblock {A simple stochastic model for the dynamics of condensation}.
\newblock {\em Journal of Physics A: Mathematical and General}, 31(1):L19, 1998.

\bibitem{evansBrazil}
M.~Evans.
\newblock Phase transitions in one-dimensional nonequilibrium systems.
\newblock {\em Brazilian Journal of Physics}, vol.30, 2000.

\bibitem{Evans2005}
M.~R. Evans and T.~Hanney.
\newblock {Nonequilibrium statistical mechanics of the zero-range process and
  related models}.
\newblock {\em Journal of Physics A: Mathematical and General}, 38(19):R195, 2005.

\bibitem{godreche}
C.~Godr{\`e}che.
\newblock {From urn models to zero-range processes: statics and dynamics}.
\newblock {\em Lecture Notes in Physics}, 716:261--294, 2007.

\bibitem{godreche2012condensation}
C.~Godr{\`e}che and J.-M. Luck.
\newblock Condensation in the inhomogeneous zero-range process: an interplay
  between interaction and diffusion disorder.
\newblock {\em Journal of Statistical Mechanics: Theory and Experiment},
  2012(12):P12013, 2012.

\bibitem{eggerssand}
J.~Eggers.
\newblock Sand as maxwell's demon.
\newblock {\em Physical Review Letters}, 83, 5322, 2009.

\bibitem{macroZRP}
Z.~Burda, D.~Johnston, J.~Jurkiewicz, M.~Kami{\'n}ski, M.~A. Nowak, G.~Papp,
  and I.~Zahed.
\newblock Wealth condensation in pareto macroeconomies.
\newblock {\em Physical Review E}, 65(2):026102, 2002.

\bibitem{trafficZRP}
D.~Chowdhury, L.~Santen, and A.~Schadschneider.
\newblock Statistical physics of vehicular traffic and some related systems.
\newblock {\em Physics Reports}, 329(4):199--329, 2000.

\bibitem{Chleboun2014}
P.~Chleboun and S.~Grosskinsky.
\newblock Condensation in stochastic particle systems with stationary product
  measures.
\newblock {\em Journal of Statistical Physics}, 154(1-2):432--465, 2014.

\bibitem{Hollander}
F.~Den~Hollander.
\newblock {\em Large deviations}, volume~14.
\newblock American Mathematical Society, 2008.

\bibitem{Touchette2009a}
H.~Touchette.
\newblock {The large deviation approach to statistical mechanics}.
\newblock {\em Physics Reports}, 478:1--95, 2009.

\bibitem{BodADD}
T.~Bodineau and B.~Derrida.
\newblock Current fluctuations in nonequilibrium diffusive systems: An
  additivity principle.
\newblock {\em Physical Review Letters} 92, 180601, 2004.

\bibitem{Giardina2011}
C.~Giardina, J.~Kurchan, V.~Lecomte, and J.~Tailleur.
\newblock {Simulating Rare Events in Dynamical Processes}.
\newblock {\em Journal of Statistical Physics}, 145:787--811, 2011.

\bibitem{Chetrite2014}
R.~Chetrite and H.~Touchette.
\newblock Nonequilibrium markov processes conditioned on large deviations.
\newblock {\em Annales Henri Poincar{\'e}}, 16(9):2005--2057, 2015.

\bibitem{angeletti2016}
F.~Angeletti and H.~Touchette.
\newblock Diffusions conditioned on occupation measures.
\newblock {\em Journal of Mathematical Physics}, 57, 023303, 2016.

\bibitem{warbook}
S.~Grosskinsky.
\newblock Interacting stochastic particle systems.
\newblock {\em London Mathematical Society Lecture Note Series},
  1(408):125--209, 2013.

\bibitem{gss}
S.~Grosskinsky, G.~M. Sch{\"u}tz, and H.~Spohn.
\newblock Condensation in the zero range process: stationary and dynamical
  properties.
\newblock {\em Journal of Statistical Physics}, 113(3-4):389--410, 2003.

\bibitem{al1}
I.~Armend{\'a}riz and M.~Loulakis.
\newblock Thermodynamic limit for the invariant measures in supercritical zero
  range processes.
\newblock {\em Probability Theory and Related Fields}, 145(1-2):175--188, 2009.

\bibitem{Landim}
C.~Landim and C.~Kipnis.
\newblock {\em {Scaling Limits of Interacting Particle Systems}}.
\newblock Springer, Berlin, 1999.

\bibitem{stamatakis}
M.~G. Stamatakis.
\newblock Hydrodynamic limit of mean zero condensing zero range processes with
  sub-critical initial profiles.
\newblock {\em Journal of Statistical Physics}, 158:87–104, 2015.

\bibitem{Schutz2007}
G.~M. Sch{\"{u}}tz and R.~J. Harris.
\newblock {Hydrodynamics of the zero-range process in the condensation regime}.
\newblock {\em Journal of Statistical Physics}, 127(2):419--430, 2007.

\bibitem{laxbook}
P.~D. Lax.
\newblock {\em Hyperbolic Systems of Conservation Laws and the Mathematical
  Theory of Shock Waves}.
\newblock SIAM, 1973.

\bibitem{paulthesis}
P.~Chleboun.
\newblock {\em {Large deviations and metastability in condensing particle
  systems}}.
\newblock PhD thesis, 2011.

\bibitem{gc2015}
S.~Grosskinsky and P.~Chleboun.
\newblock A dynamical transition and metastability in a size-dependent
  zero-range process.
\newblock {\em Journal of Physics A: Mathematical and Theoretical} 48 (5),
  055001, 2015.

\bibitem{Touchette2014}
H.~Touchette.
\newblock Equivalence and nonequivalence of ensembles: Thermodynamic,
  macrostate, and measure levels.
\newblock {\em Journal of Statistical Physics}, 159(5):987--1016, 2015.

\bibitem{CGfinite}
S.~Grosskinsky and P.~Chleboun.
\newblock Finite size effects and metastability in zero-range condensation.
\newblock {\em Journal of Statistical Physics}, 140: 846–872, 2010.

\bibitem{transcomplex}
A.~Schadschneider, D.~Chowdhury, and K.~Nishinari
\newblock {\em Stochastic Transport in Complex Systems: from Molecules to Vehicles}.
\newblock Elsevier, 2011.

\end{thebibliography}

\end{document}